\documentclass[useAMS,fleqn,usenatbib]{mnras}

\usepackage{newtxtext, newtxmath}
\usepackage[T1]{fontenc}
\usepackage{ae,aecompl}

\usepackage{graphicx}
\usepackage{color}
\usepackage{soul}
\usepackage{bm}

\newcommand{\cs}{c_\mathrm{s}}
\newcommand{\warp}{|\psi|}

\title[Local numerical simulations of warped discs]{Local numerical simulations of warped discs}

\author[S.-J. Paardekooper \& G.I. Ogilvie]{Sijme-Jan Paardekooper$^{1,2}$\thanks{E-mail:  s.j.paardekooper@qmul.ac.uk} and 
Gordon I. Ogilvie$^2$\\
$^1$Astronomy Unit, School of Physics and Astronomy, Queen Mary, University of London, Mile End Road, London E1 4NS\\
$^2$DAMTP, University of Cambridge, Wilberforce Road, Cambridge CB3 0WA
}

\date{Draft version \today}
\pubyear{2018}

\begin{document}
\label{firstpage}
\pagerange{\pageref{firstpage}--\pageref{lastpage}}
\maketitle


\begin{abstract}
We study the hydrodynamical stability of the laminar flows associated with warped astrophysical discs using numerical simulations of warped shearing boxes. We recover linear growth rates reported previously due to a parametric resonance of inertial waves, and show that the nonlinear saturated state can significantly reduce the laminar flows, meaning that the warp would evolve on much longer time scales than would be concluded from the internal torques due to these laminar flows. Towards larger warp amplitudes, we find first of all a reversal of angular momentum flux, indicating that the mass distribution would evolve in an anti-diffusive manner, and second that the linear growth rates disappear, possibly because of the very strong shear in the laminar flows in this regime. For discs with small enough viscosity, a nonlinear state can still be found when linear growth rates are absent by introducing a large enough perturbation, either by starting from a nonlinear state obtained at smaller warp amplitude, or by starting from a state with no laminar flows.
\end{abstract}
 


\section{Introduction}
\label{secIntro}

Discs are ubiquitous in astrophysics and can be found for example around young stars, supermassive black holes, in cataclysmic variables and around planets, most famously Saturn. While in the simplest picture a disc has a single orbital plane, the orbital plane can vary with radius if there is some form of misalignment present in the system. In that case, the disc is said to be \emph{warped}.  

The misalignment necessary to induce a warp in a disc can for example come from accreted material that has a different angular momentum vector \citep{bate10}, a distant companion on an inclined orbit \citep{papaloizou95} or a misaligned magnetic field \citep{lai99}. Warps can also be excited through radiation \citep{pringle96}, winds \citep{schandl94} or tidal effects \citep{lubow92}. The classic case of misalignment is that of an accretion disc for which the orbital axis does not coincide with the spin axis of the central black hole \citep{bardeen75}. Observationally, we know of warped discs in X-ray binaries \citep[e.g.][]{katz73, kotze12}, and around black holes in galaxy centres \citep[e.g.][]{miyoshi95,greenhill05}. Understanding the evolution of such warped discs due to internal and external torques is a rich and intricate problem.

The first consistent linear theory of the evolution of a warped viscous Keplerian disc was provided by \cite{papaloizou83}, who improved upon earlier work \citep[e.g.][]{bardeen75, petterson77, petterson78,hatchett81} by including the correct form of internal torques necessary to conserve angular momentum. Parametrising the viscosity through the $\alpha$-parameter \citep{shakura73}, they found that the warp diffuses on a time scale that is a factor $\alpha^2$ shorter than the viscous time scale in the regime $H/r < \alpha \ll 1$ (the diffusive regime), where $H$ is the disc angular semithickness. A transition to wavelike propagation of the warp occurs when $\alpha <  H/r$ \citep{papaloizoulin95}. A fully nonlinear theory for the diffusive regime in Keplerian discs (and for bending waves in non-Keplerian discs) was developed in \cite{ogilvie99}. This analysis also provided a means to calculate the internal torques governing the evolution of the warp from the amplitude of the warp and other parameters. 

The internal fluid motions in a warped disc, driven by a horizontal pressure gradient stemming from the change of the orbital plane with radius, are usually treated as laminar flows. It has been noted that these oscillatory shear flows could in fact be hydrodynamically unstable \citep{papaloizou95, gammie00}. The resulting state may be fully turbulent but will at least involve significant wave activity altering the laminar flows and therefore the internal torques \citep{OL13b}. This means one has to rely on numerical simulations to determine these torques and therefore the evolution of warped discs.  

Global numerical hydrodynamical simulations of warped accretion discs are computationally demanding for several reasons. As is usual for astrophysical discs, the range of time scales and length scales is substantial. For warped discs in particular, the calculations are in addition necessarily three-dimensional. Moreover, the vertical extent of the computational domain has to be significantly larger compared to unwarped discs to allow the orbital plane to vary with radius. This leads to regions in the domain of extremely low density, which poses a challenge to most grid-based codes. Most simulations of warped discs have been done using smoothed particle hydrodynamics (SPH) \citep[e.g.][]{nelson99, lodato07, lodato10, nixon12, xiang13, nealon16}. Grid-based global calculations were done by e.g. \cite{fragile07, fragner10, krolik15}. 

In none of the global simulations mentioned above was a hydrodynamic instability seen, perhaps due to a lack of resolution. A local model of a warp, harnessing the power of the shearing box formalism \citep[e.g.][]{goldreich65} to deal with warped astrophysical discs, was developed in \cite{OL13a}. The warped shearing box has been used recently by \cite{paris18} to examine the effect of a magnetic field on the internal dynamics of a warped disc and on the propagation of a warp.  An analogous model has also been developed by \cite{ogilvie14} for eccentric discs. \cite{barker14} used this model to analyse the parametric instability of inertial waves in an eccentric disc.  More recently, \cite{wienkers18} have followed the nonlinear evolution of the instability in 2D numerical simulations in an eccentric shearing box.  Breaking of the waves away from the midplane was found to limit the growth of unstable modes, and a competitive dynamics ensued between the inertial waves and zonal flows that developed in the box.  While the resulting Reynolds stresses were not very effective in transporting angular momentum, they would have the effect of damping the eccentricity.

In the local model that is the warped shearing box it is easier to obtain high resolution, while the potentially unstable laminar flows are still present. The hydrodynamic laminar flows were computed in \cite{OL13a}, while their stability was investigated in \cite{OL13b}. They found widespread hydrodynamic instability where a pair of inertial waves can grow exponentially by coupling to the laminar warp motion \citep[see also][]{gammie00}. As it is the nonlinear outcome of this instability that in the end will determine the internal torques, numerical simulations are needed to calculate these. This is the subject of this paper. 

We start in section \ref{secEq} with the governing equations and the definitions of the components of the internal torque that we are aiming to measure. In section \ref{secNum} we present the numerical method, which is tested in section \ref{secTest}. Results are presented in section \ref{secRes} and we conclude in section \ref{secDisc}. 


\section{Basic equations}
\label{secEq}

We will be using the local, isothermal model of a warped disc as derived in \cite{OL13a}. Starting with the standard shearing box equations, a transformation is made to warped shearing coordinates that follow the warped orbital motion\footnote{In \cite{OL13a} the warped shearing coordinates are denoted with primes. In this paper, we work exclusively with warped shearing coordinates and therefore we omit the primes for clarity.}. In this model, the warp is represented by a dimensionless amplitude $\warp$ and the governing equations read \citep[equations (38)-(40) and (42) in][]{OL13a}:
\begin{eqnarray}
\partial_t \rho+\partial_x(\rho v_x)+\partial_y[\rho(v_y+q\tau v_x)]+\nonumber\\
\partial_z[\rho(v_z+\warp\cos\tau v_x)]=0,\label{eqSysFirst}\\
\partial_t v_x+v_x\partial_x v_x +(v_y+q\tau v_x)\partial_y v_x+\nonumber\\ (v_z+\warp\cos\tau v_x)\partial_z v_x+\nonumber\\ (\partial_x p+q\tau\partial_y p+\warp\cos\tau \partial_z p)/\rho = 2\Omega v_y,\\
\partial_t v_y+v_x\partial_x v_y +(v_y+q\tau v_x)\partial_y v_y+\nonumber\\ (v_z+\warp\cos\tau v_x)\partial_z v_y+\nonumber\\ \partial_y p/\rho = (q-2)\Omega v_x,\\
\partial_t v_z+v_x\partial_x v_z +(v_y+q\tau v_x)\partial_y v_z+\nonumber\\(v_z+\warp\cos\tau v_x)\partial_z v_z+\nonumber\\\partial_z p/\rho = -\warp\Omega\sin\tau v_x -\Omega^2z.\label{eqSysLast}
\end{eqnarray}
Here, $q$ is the dimensionless rate of orbital shear ($q=1.5$ in a Keplerian disc), $\Omega$ the orbital angular velocity of the box,  $\tau=\Omega t$ the orbital phase, $\rho$ is the fluid density, ${\bf v}=(v_x, v_y, v_z)^T$ is the velocity relative to the warped orbital motion. It is related to the standard velocity in the rotating frame of the shearing box ${\bf u}$ by:
\begin{eqnarray}
v_x &=& u_x,\\
v_y &=& u_y + q\Omega x,\\
v_z &=& u_z - \warp\Omega_0\sin\tau x.
\end{eqnarray}
Note that throughout this paper the components of the velocity vector (and, later, of the viscous stress tensor) are referred to the standard orthonormal Cartesian basis.

We use an isothermal equation of state relating pressure and density:
\begin{equation}
p = \cs^2\rho,
\end{equation}
with $\cs$ the isothermal sound speed. For brevity, we have omitted viscous forces. The three dimensionless numbers characterising a particular physical setup are the warp amplitude $\warp$, the shear parameter $q$ and the (shear) viscosity $\alpha$ (see equation (\ref{eq:alpha})). Note that the aspect ratio $H/r = \cs/(r\Omega)$ is \emph{not} one of the dimensionless parameters of the local model. 

In the absence of a warp ($\warp=0$), $v_x=v_y=v_z=0$ is a solution, with a density profile
\begin{equation}
\rho=\frac{\Sigma}{\sqrt{2\pi}}\exp\left(-\frac{\Omega^2z^2}{2\cs^2}\right),
\end{equation}
where $\Sigma$ is the surface density. A warp introduces a laminar flow in the disc, which in the local model only depends on $z$ and $\tau$ \citep{OL13a}. A nonlinear separation of variables:
\begin{eqnarray}
v_x(z, t) &=& u(\tau)\Omega z,\\
v_y(z, t) &=& v(\tau)\Omega z,\\
v_z(z, t) &=& w(\tau)\Omega z,\\
h(z,t) &=& \cs^2 f(\tau) - \frac{1}{2}\Omega^2z^2 g(\tau),
\end{eqnarray}
where $h=\cs^2\log \rho + \mathrm{constant}$ is the pseudo-enthalpy, leads to a set of ordinary differential equations for the dimensionless amplitudes $u$, $v$, $w$, $f$ and $g$ \citep{OL13a}. These amplitudes are periodic in $\tau$ with period $2\pi$ and accurate solutions can be obtained by standard numerical methods for solving ordinary differential equations.

The evolution of both warped and unwarped discs is governed by transport of angular momentum, and therefore by an internal torque $\bm{\mathcal{G}}$. In a warped disc, there are three relevant components of $\bm{\mathcal{G}}$, with dimensionless coefficients \citep{ogilvie99, OL13a}
\begin{eqnarray}
Q_1 &=& -\frac{\left<\int (\rho v_x v_y - T_{xy})dz\right>_{h,\tau}}{\Sigma \cs^2},\\
Q_2 &=& \frac{\left<\cos\tau\int \rho v_x\Omega z dz- \sin\tau\int (\rho v_x v_z - T_{xz})dz\right>_{h,\tau}}{\warp\Sigma \cs^2},\\
Q_3 &=& \frac{\left<\cos\tau\int (\rho v_x v_z - T_{xz})dz + \sin\tau\int \rho v_x\Omega zdz\right>_{h,\tau}}{\warp\Sigma \cs^2},
\end{eqnarray}
where $T_{ij}$ are the components of the viscous stress tensor and the subscript $h,\tau$ indicates that a horizontal and time average over one period is taken. The vertical component of the torque, $Q_1$, involves the radial transport of vertical angular momentum and is similar to the usual torque in an unwarped accretion disc. If negative, it causes the mass distribution to evolve diffusively. The horizontal components of the torque, $Q_2$ and $Q_3$, involve the radial transport of horizontal angular momentum. If $Q_2>0$, it causes a diffusion of the warp, while $Q_3$ causes a dispersive wavelike propagation of the warp. 

While the torque components can be calculated for the laminar flow induced by the warp \citep{OL13a}, it was found that these flows may in fact be hydrodynamically unstable \citep{OL13b}. Therefore, in order to measure the torque components in the nonlinear state of the instability one has to resort to numerical simulations. This is the purpose of this paper. 


\section{Numerical method}
\label{secNum}

In order to use a finite volume volume numerical scheme we write the system (\ref{eqSysFirst})-(\ref{eqSysLast}) in conservative form:
\begin{eqnarray}
\partial_t \rho+\partial_x(\rho v_x)+\partial_y[\rho(v_y+q\tau v_x)]+\nonumber\\
\partial_z[\rho(v_z+\warp\cos\tau v_x)]=0,\\
\partial_t (\rho v_x)+\partial_x(\rho v_x^2+p) +\partial_y [\rho v_x(v_y+q\tau v_x)+q\tau p]+\nonumber\\\partial_z [\rho v_x(v_z+\warp\cos\tau v_x)+\warp\cos\tau p] = 2\Omega \rho v_y,\\
\partial_t (\rho v_y)+\partial_x(\rho v_x v_y) +\partial_y[\rho v_y(v_y+q\tau v_x)+p]+\nonumber\\\partial_z[\rho v_y(v_z+\warp\cos\tau v_x)]= (q-2)\Omega \rho v_x,\\
\partial_t (\rho v_z)+\partial_x(\rho v_x v_z) +\partial_y[\rho v_z(v_y+q\tau v_x)] +\nonumber\\\partial_z[\rho v_z(v_z+\warp\cos\tau v_x)+p]= -\warp\Omega\sin\tau \rho v_x -\Omega^2z\rho.
\end{eqnarray}
Note that, in the absence of a warp ($\warp=0$), this system reduces to the standard equations for a shearing box in shearing coordinates, and that it can be written as
\begin{equation}
\partial_t {\bf W} + \partial_x {\bf F}_x + \partial_y {\bf F}_y + \partial_z {\bf F}_z = {\bf S},
\end{equation}
where ${\bf W}$ is the state vector, ${\bf F}_{xyz}$ are the flux vectors in three directions and ${\bf S}$ is the source vector. Then, we use the technique of operator splitting to treat each spatial dimension separately. 


\subsection{Dimensional splitting}


\subsubsection{$x$-direction}
For the $x$-direction, we keep only the flux term with ${\bf F}_x$ and the associated source terms \citep[see][]{eulderink95}:
\begin{eqnarray}
\partial_t \rho+\partial_x(\rho v_x)=0,\\
\partial_t (\rho v_x)+\partial_x(\rho v_x^2+p) = 2\Omega \rho v_y,\\
\partial_t (\rho v_y)+\partial_x(\rho v_x v_y)= (q-2)\Omega \rho v_x,\\
\partial_t (\rho v_z)+\partial_x(\rho v_x v_z)= -\warp\Omega\sin\tau \rho v_x.
\end{eqnarray}
Source terms are integrated using a Crank-Nicholson scheme allowing for exact integration of epicyclic motion \citep{stone10}. These equations are similar to the ordinary shearing box equations but for the extra source term $\propto \warp$, and we refer to \cite{paardekooper12} for details of the implementation. 


\subsubsection{$y$-direction}

In this paper, we only consider ``axisymmetric" shearing boxes, i.e. no quantity depends on $y$. This situation corresponds to studying warped discs in which the azimuthal variation is on a global scale comparable to the circumference of the disc, rather than on the local length scale $H=\cs/\Omega$. For completeness, the equations for the $y$-direction read: 
\begin{eqnarray}
\partial_t \rho+\partial_y[\rho(v_y+q\tau v_x)]=0,\\
\partial_t (\rho v_x)+\partial_y [\rho v_x(v_y+q\tau v_x)+q\tau p]=0,\\
\partial_t (\rho v_y)+\partial_y[\rho v_y(v_y+q\tau v_x)+p]=0,\\
\partial_t (\rho v_z)+\partial_y[\rho v_z(v_y+q\tau v_x)] =0.
\end{eqnarray}

Note that for $q\tau \gg 1$ the system will be completely dominated by the shearing of the coordinate frame. Therefore, it is necessary to occasionally remap the solution to $\tau=0$, by shifting the solution in the $y$-direction by an amount $q x\tau_\mathrm{remap}$. For a carefully chosen $\tau_\mathrm{remap}$, this can be done without interpolation. Basically, if the grid is symmetric with respect to $x=0$, and the row with smallest $|x|$ has to be shifted by 1 grid cell (which can be done without interpolation), then the next row will have to be shifted by two cells (again without interpolation), etc. 
 

\subsubsection{$z$-direction}

For the $z$-direction, we keep only the flux term with ${\bf F}_z$ and the corresponding source term:
\begin{eqnarray}
\partial_t \rho+\partial_z[\rho(v_z+\warp\cos\tau v_x)]=0,\\
\partial_t (\rho v_x)+\partial_z [\rho v_x(v_z+\warp\cos\tau v_x)+\warp\cos\tau p] = 0,\\
\partial_t (\rho v_y)+\partial_z[\rho v_y(v_z+\warp\cos\tau v_x)]= 0,\\
\partial_t (\rho v_z)+\partial_z[\rho v_z(v_z+\warp\cos\tau v_x)+p]= -\Omega^2z\rho.
\end{eqnarray}
Fluxes now contain terms $\propto \warp\cos\tau$, which find their way into the eigenvalues of the Jacobian matrix $\mathsf{A} = d{\bf F}_z/d{\bf W}$ through functions $\xi\equiv \warp\cos\tau$ and $\eta \equiv\sqrt{1+\xi^2}$:
\begin{eqnarray}
\lambda_1 &=& v_z+\xi v_x - \eta \cs,\\
\lambda_2 &=& v_z+\xi v_x + \eta \cs,\\
\lambda_3 &=& v_z+\xi v_x,\\ 
\lambda_4 &=& v_z+\xi v_x,
\end{eqnarray}
and the corresponding eigenvectors:
\begin{eqnarray}
{\bf e}_1 &=& (1, v_z - \cs/\eta, v_x - \xi \cs/\eta, v_y)^T,\\
{\bf e}_2 &=& (1, v_z + \cs/\eta, v_x+\xi \cs/\eta, v_y)^T,\\
{\bf e}_3 &=& (0, \xi, 1, 0)^T,\\
{\bf e}_4 &=& (0, 0, 0, 1)^T.
\end{eqnarray}
A flux difference at the interface between cells $i$ and $i-1$ is projected onto the eigenvectors of a suitably averaged Jacobian $\hat{\mathsf{A}}$:
\begin{equation}
{\bf F}_i - {\bf F}_{i-1}= \sum_k \hat \lambda_k a_k \hat{{\bf e}}_k 
\end{equation}
 with projection coefficients:
\begin{eqnarray}
a_1 &=& -\frac{\Delta_z - (\hat{v}_z+ \xi\hat{v}_x+ \eta \cs)\Delta_\rho+\xi\Delta_x}{2\eta \cs},\\
a_2 &=& \frac{\Delta_z - (\hat{v}_z+\xi\hat{v}_x - \eta \cs)\Delta_\rho+\xi\Delta_x}{2\eta \cs},\\
a_3 &=& \frac{\Delta_x - \left(\hat{v}_x-\xi \hat{v}_z\right)\Delta_\rho - \xi\Delta_z}{\eta^2},\\
a_4 &=& \Delta_y - \hat{v}_y\Delta_\rho,
\end{eqnarray}
where ${\bf F}_i - {\bf F}_{i-1} = (\Delta_\rho,\Delta_x,\Delta_y,\Delta_z)^T$ and $\hat{{\bf v}}$ denotes the Roe-averaged velocity. The first order interface flux is then given by 
\begin{equation}
{\bf F}_{i-1/2} = \frac{1}{2}\left({\bf F}_{i-1}+{\bf F}_i - \sum_k |\hat\lambda_k|a_k \hat{\bf e}_k\right).
 \end{equation}
Second order corrections are obtained using a flux limiter \citep{paardekooper12}.

The dependence on $\tau$ is periodic, so there is no need to remap if simulating an axisymmetric disc (i.e. no $y$-dependence). Otherwise it is advantageous to pick $\tau_\mathrm{remap}$ a multiple of $2\pi$. In practice, this can be achieved by slightly modifying the extent of the domain in $x$ and $y$.  


\subsection{Viscosity}
\label{secVisc}


Viscous source terms are integrated separately from the hyperbolic part of the governing equations using the technique of operator splitting. In this step, we solve for the change in velocities due to viscous forces only, which in warped shearing coordinates take the form:
\begin{eqnarray}
\frac{\partial}{\partial t}(\rho v_x) &=& \frac{\partial T_{xx}}{\partial x} +  q\tau \frac{\partial T_{xx}}{\partial y} + \warp\cos\tau \frac{\partial T_{xx}}{\partial z} +\nonumber\\
& &\frac{\partial T_{xy}}{\partial y} +\frac{\partial T_{xz}}{\partial z}\\
\frac{\partial}{\partial t}(\rho v_y) &=& \frac{\partial T_{xy}}{\partial x} +  q\tau \frac{\partial T_{xy}}{\partial y} + \warp\cos\tau \frac{\partial T_{xy}}{\partial z} +\nonumber\\
& &\frac{\partial T_{yy}}{\partial y} +\frac{\partial T_{yz}}{\partial z}\\
\frac{\partial}{\partial t}(\rho v_z) &=& \frac{\partial T_{xz}}{\partial x} +  q\tau \frac{\partial T_{xz}}{\partial y} + \warp\cos\tau \frac{\partial T_{xz}}{\partial z} +\nonumber\\
& &\frac{\partial T_{yz}}{\partial y} +\frac{\partial T_{zz}}{\partial z},
\end{eqnarray}
where $T_{ij}$ denote the components of the viscous stress tensor, which, neglecting bulk viscosity, read:
\begin{eqnarray}
T_{xx} &=&  2\rho \nu \left(\frac{\partial v_x}{\partial x} + q\tau \frac{\partial v_x}{\partial y} + \warp\cos\tau \frac{\partial v_x}{\partial z}-\frac{D}{3}\right) \\
T_{xy} &=&  \rho \nu \left(\frac{\partial v_x}{\partial y} + \frac{\partial v_y}{\partial x} - q\Omega + q\tau \frac{\partial v_y}{\partial y} + \warp\cos\tau\frac{\partial v_y}{\partial z}\right) \\
T_{xz} &=&  \rho \nu \left(\frac{\partial v_x}{\partial z} + \frac{\partial v_z}{\partial x} + \warp\sin\tau+ q\tau \frac{\partial v_z}{\partial y} + \warp\cos\tau\frac{\partial v_z}{\partial z}\right)  \\
T_{yy}  &=&  2\rho \nu \left(\frac{\partial v_y}{\partial y} -\frac{D}{3}\right) \\
T_{yz}  &=&  \rho \nu \left(\frac{\partial v_y}{\partial z} + \frac{\partial v_z}{\partial y} \right)  \\
T_{zz} &=&  2\rho \nu \left(\frac{\partial v_z}{\partial z} -\frac{D}{3}\right),
\end{eqnarray}
where the velocity divergence
\begin{equation}
D = \frac{\partial v_x}{\partial x} + q\tau \frac{\partial v_x}{\partial y} + \warp\cos\tau \frac{\partial v_x}{\partial z} + \frac{\partial v_y}{\partial y} + \frac{\partial v_z}{\partial z}.
\end{equation}
The constant shear viscosity $\nu$ is parameterised in the usual way as 
\begin{equation}
\nu=\alpha \cs^2/\Omega.
\label{eq:alpha}
\end{equation}
The viscous terms are integrated using a simple explicit Euler step, which was found to be accurate enough in the regime of interest.  
  

\begin{figure}
\includegraphics[width=\columnwidth]{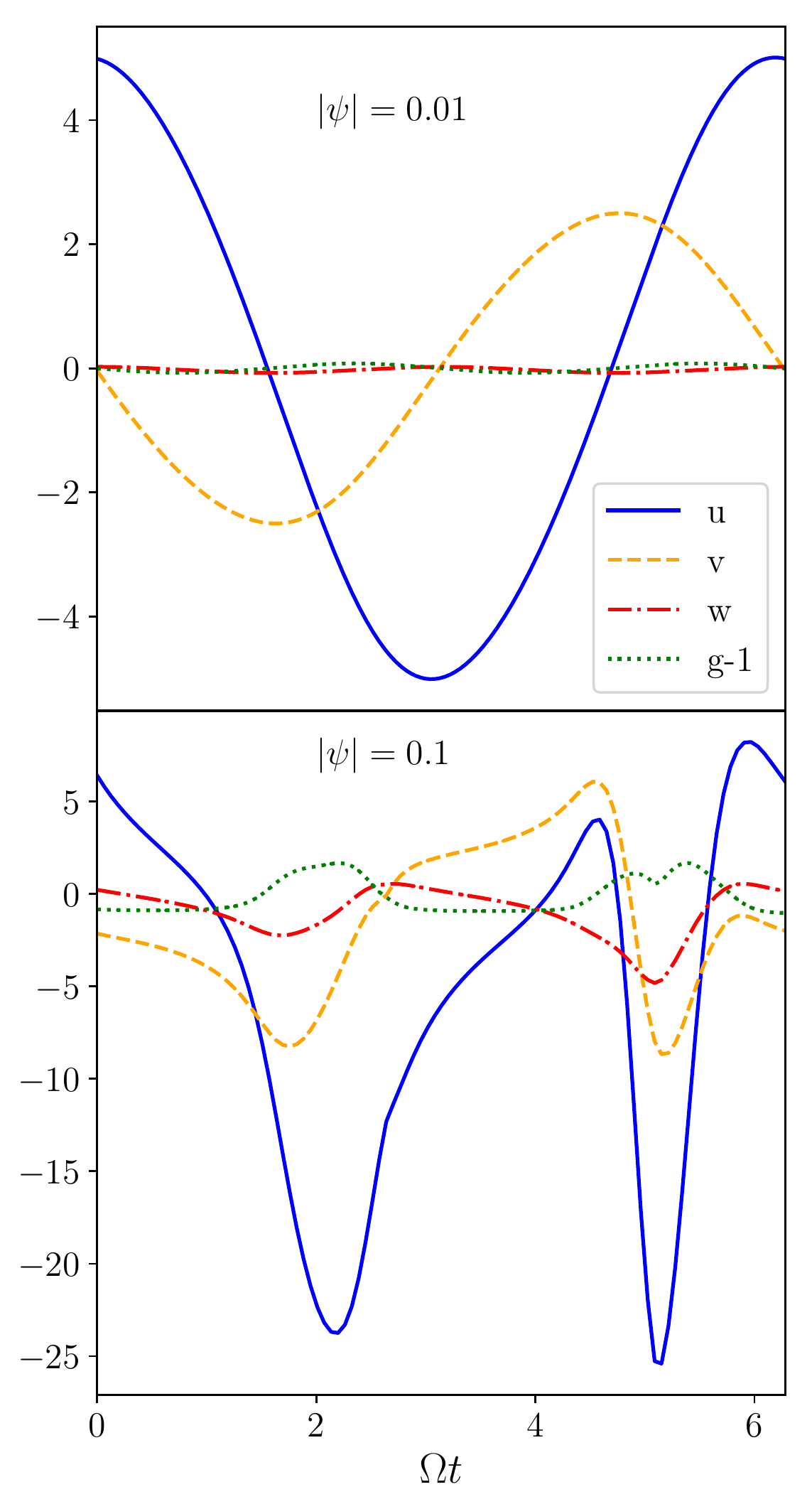}
\caption{Laminar flow solutions for $q=1.5$ and $\alpha=0.001$. Top panel: $\warp=0.01$, bottom panel: $\warp=0.1$.}
\label{fig:lam_sh15}
\end{figure}

\section{Initial conditions and tests}
\label{secTest}

In units of $H=\cs/\Omega$, we take the computational domain to be within $-6 < z/H < 6$. The horizontal extent of the computational domain is such that it can fit an integer number of inertial waves of interest (see Section \ref{secInertial} below), but typically $-4 < x/H < 4$. This domain is covered by a uniform mesh of at least $(N_x, N_z)=(128, 192)$, so that a scale height $H$ is resolved by at least $16$ cells. In the figures below, we will refer to this base resolution as ``$N=128$", and ``$N=256$" refers to $(N_x, N_z)=(256, 384)$, etc. The boundaries are periodic in $x$, while for the vertical boundaries we use non-reflecting conditions \citep[see][]{paardekooper06}.


\subsection{Laminar flows}
\label{secLaminar}

The laminar flows discussed in section \ref{secEq} provide a first basic test for the implementation of the Roe solver for the warped shearing box. For small $\warp$ in a Keplerian $q=1.5$ disc, one expects $u,v \propto \warp/\alpha$ and to be sinusoidal, with $w$ and $g-1$ a factor $\warp$ smaller \citep{OL13a}. Such a solution is presented in the top panel of figure \ref{fig:lam_sh15}. Note that even for such a small warp amplitude $\warp=0.01$ the horizontal flow speeds exceed the sound speed sufficiently high above the mid plane of the disc (since for example $v_x = u(\tau)\Omega z$), which is due to the fact that the Keplerian case suffers from a resonance due to the coincidence of the orbital and epicyclic frequency. This resonance is moderated by viscosity, but since typically $\alpha \ll 1$ the resulting horizontal velocities can be very high. For larger warp amplitudes, the laminar flows become less sinusoidal and more extreme, as illustrated in the bottom panel of figure \ref{fig:lam_sh15}.

\begin{figure}
\includegraphics[width=\columnwidth]{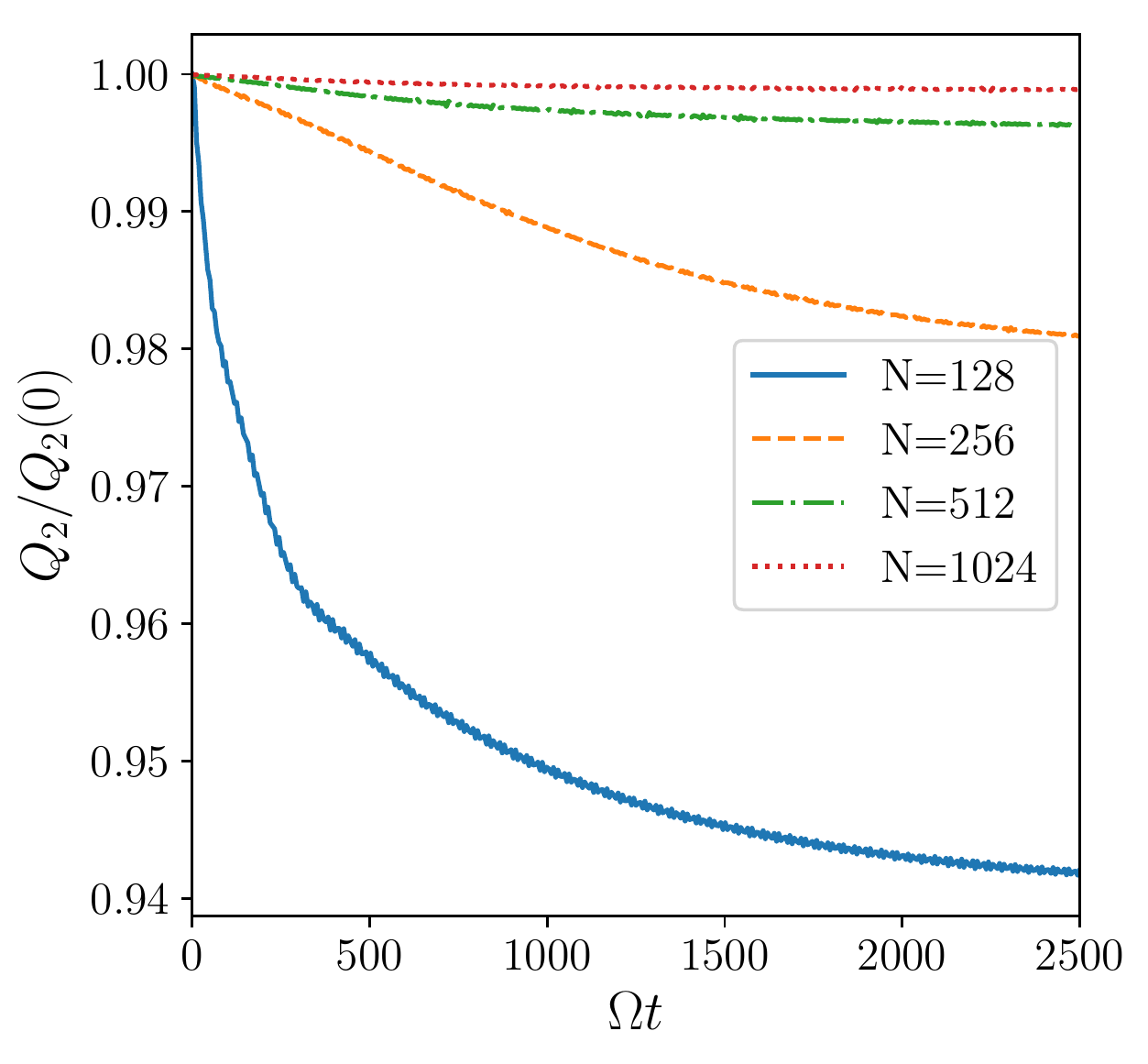}
\caption{Evolution of the $Q_2$ torque component, relative to its initial value, for $\warp=0.01$, $\alpha=0.001$ and $q=1.5$, where all quantities depend only on $z$ and $t$ so that the laminar warp motion should be stable. }
\label{fig:lam_evolve_psi-2}
\end{figure}

The laminar solutions obtained with the hydrodynamic solver compare well with those obtained using a standard ordinary differential equation solver. More challenging is to follow the laminar flow accurately for many (hundreds to thousands) periods, which is necessary when the growth rate of any instability feeding off the laminar flow is very small. Any evolution in the laminar flow can be conveniently observed in the torque component $Q_2$, which measures roughly speaking the diffusion of the warp \citep{OL13a}. For the purely laminar flow one expects \citep{OL13a}:
\begin{equation}
Q_2 = \frac{1+7\alpha^2}{\alpha(4+\alpha^2)} + O(\warp^2).
\label{eqQ2approx}
\end{equation}
In a one dimensional calculation, where all quantities depend only on $z$ and $t$, the laminar flows should be stable and therefore $Q_2$ should be constant. In reality, numerical diffusion can act to reduce the amplitude of the laminar motions. In figure \ref{fig:lam_evolve_psi-2} we show the evolution of $Q_2$ for four different resolutions and $\warp=0.01$, $\alpha=0.001$ and $q=1.5$. At low resolution, the amplitude of the laminar motions is reduced by more than $5\%$ over $2500$ periods. This reduction is mainly due to difficulties at high altitude $(z/H > 3)$, where the densities get very low and any slight error in maintaining hydrostatic equilibrium may result in an unphysical state. The use of flux limiters can prevent this from happening at the expense of more numerical diffusion. This has the effect that the very upper regions at low resolution no longer participate in the laminar motion, resulting in a reduction of $Q_2$. A similar reduction in $Q_2$ would be obtained by choosing a smaller vertical domain. In fact, the reason for choosing the upper boundary at $z/H =6$ is based on getting the correct values of the torque components. 

The numerical diffusion of the laminar flow was found to scale with the amplitudes of the flow. For example, for $\warp=\alpha=0.01$, the amplitudes are reduced by an order of magnitude relative to the case considered above, leading to the torque $Q_2$ component to be preserved to within $0.2\%$ even at our lowest resolution $N=128$.  

\begin{figure}
\includegraphics[width=\columnwidth]{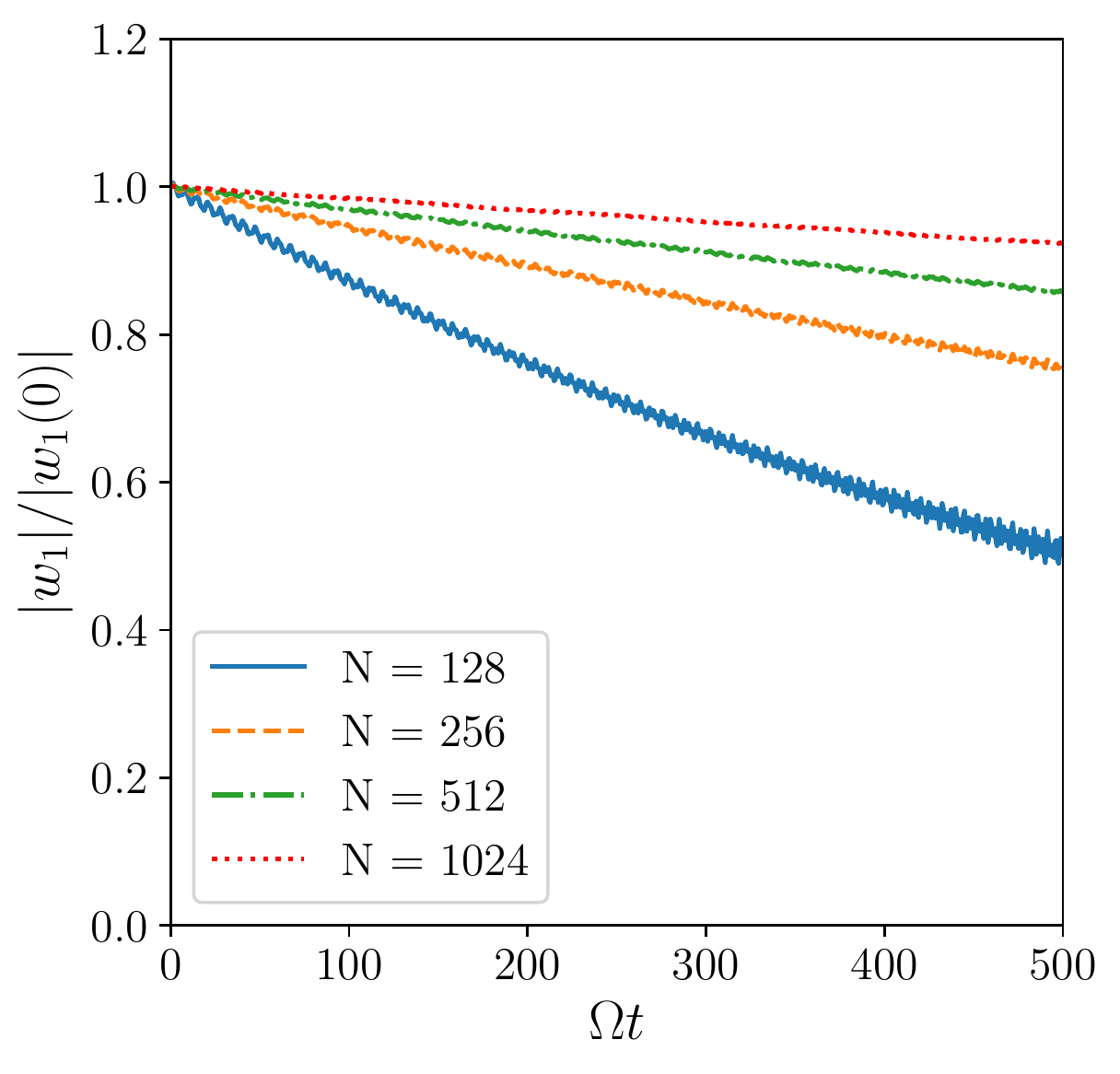}
\caption{Evolution of the $n=1$, $k_x=1.6$ Fourier-Hermite component of the vertical velocity (normalised by the value at $t=0$) for $\alpha=0$, $q=1.5$ and $\warp=0$ for different resolutions.}
\label{fig:FH_inertial}
\end{figure}

\begin{figure}
\includegraphics[width=\columnwidth]{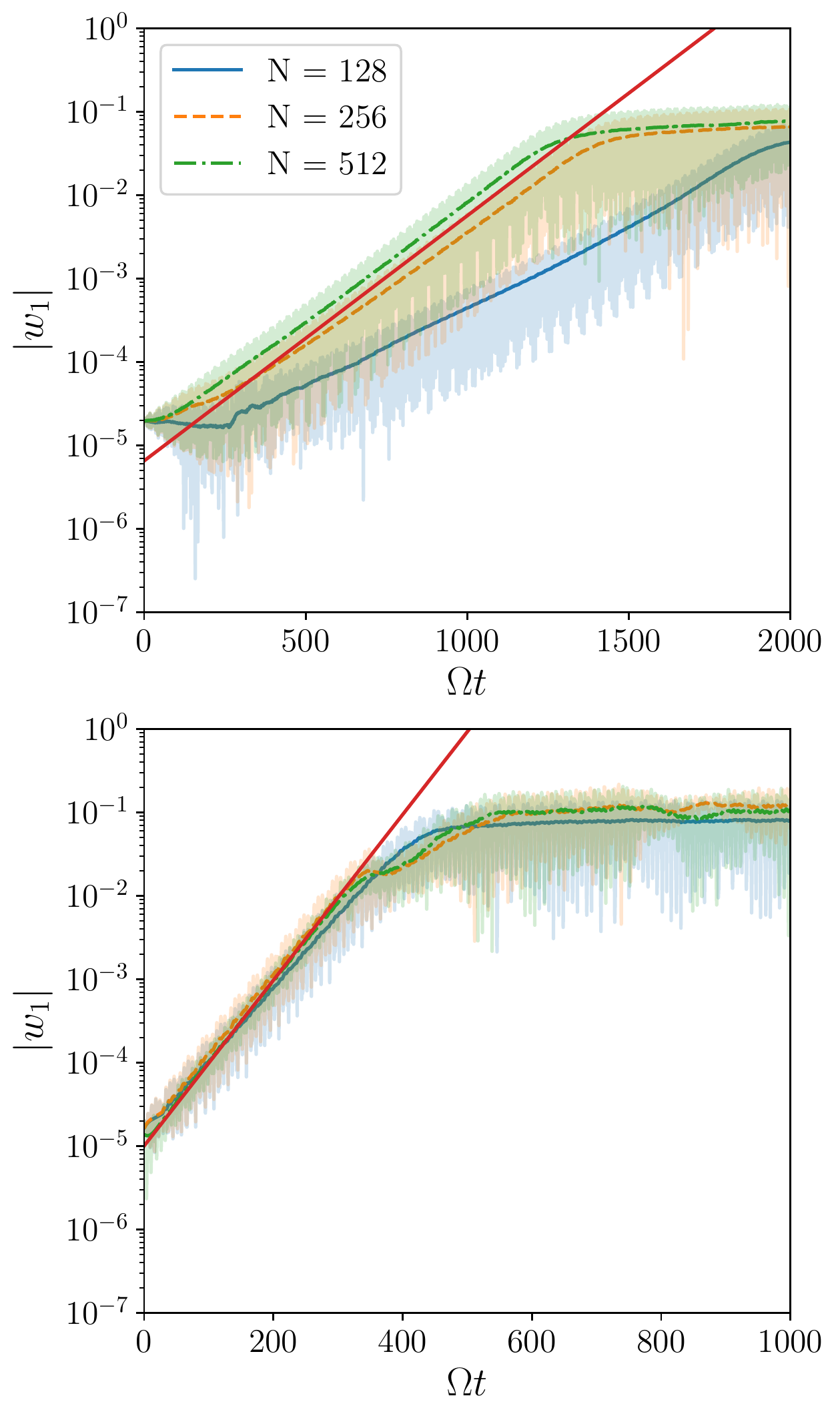}
\caption{Evolution of the magnitude of the $n=1$, $k_x = 1.6$ Fourier-Hermite component of the vertical velocity in an inviscid non-Keplerian $q=1.6$ disc with $\warp=0.01$ (top panel) and $\warp=0.04$ (bottom panel) for different resolutions. The solid red line indicates the growth rate determined from a linear calculation.}
\label{fig:FH_sh16}
\end{figure}

\begin{figure}
\includegraphics[width=\columnwidth]{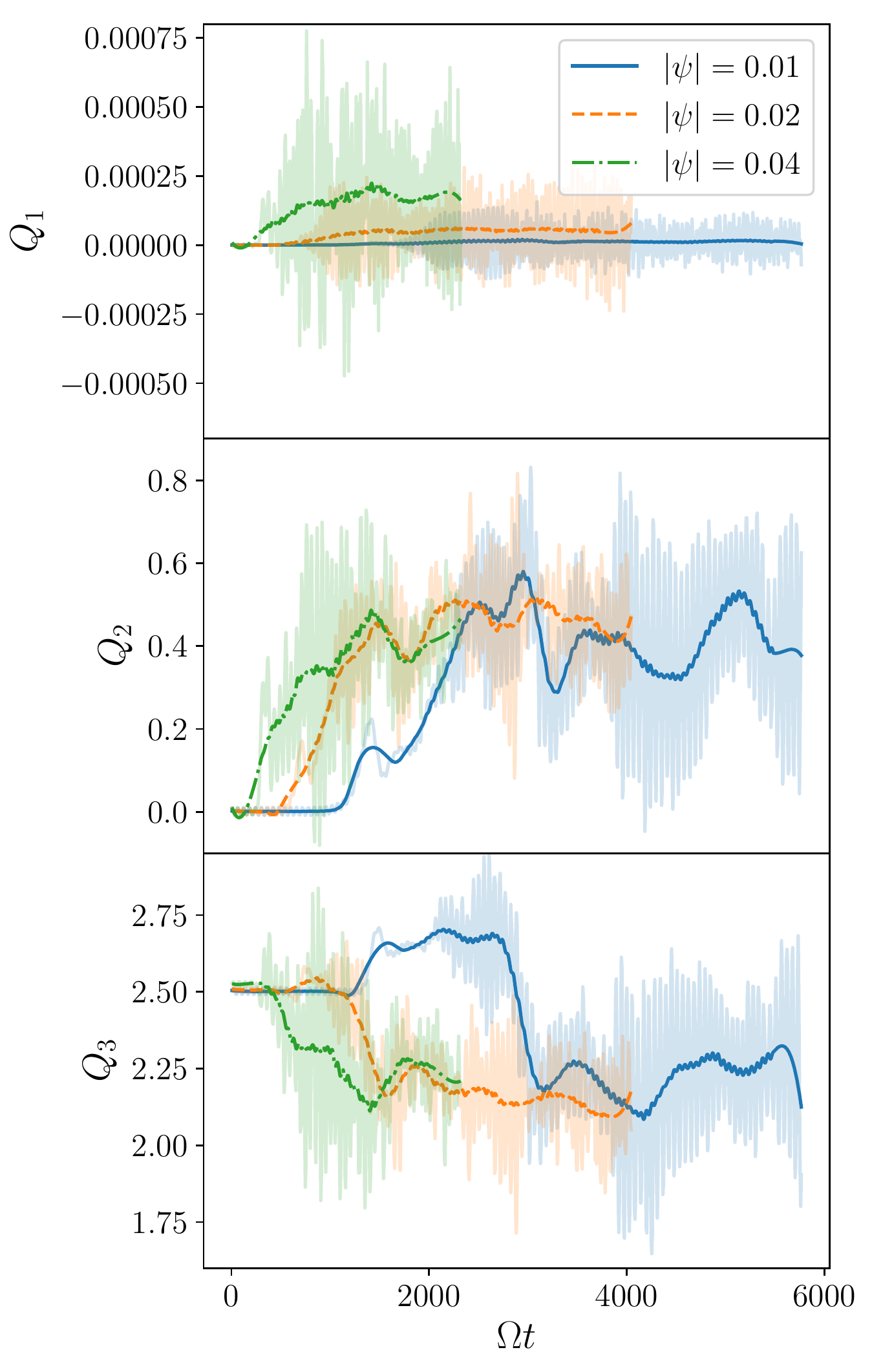}
\caption{Evolution of the torque components for an inviscid disc with $q=1.6$ for three different warp parameters, all at resolution $N=512$. The actual data are shown by the transparent curves, while the opaque curves are smoothed for readability.}
\label{fig:Q_sh16}
\end{figure}

\begin{figure*}
\includegraphics[width=\textwidth]{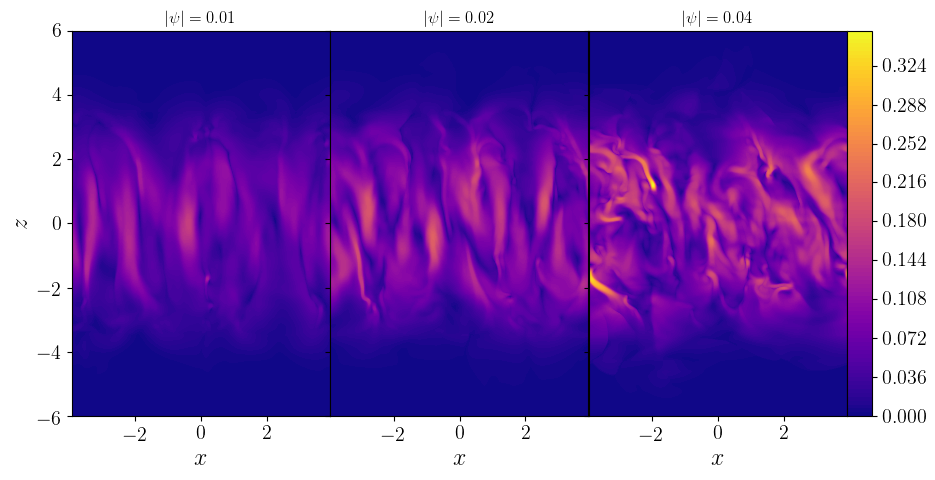}
\caption{Snapshots of $\sqrt{\rho}|{\bf \delta v}|$, which is a measure of wave energy, for an inviscid disc with $q=1.6$ in the nonlinear phase for three different warp amplitudes, all at resolution $N=512$. Left panel: $\warp=0.01$ at $\Omega t=1600\pi$. Middle panel: $\warp=0.02$ at $\Omega t = 1200\pi$. Right panel: $\warp=0.04$ at $\Omega t = 600\pi$.}
\label{fig:velx_sh16}
\end{figure*}

\subsection{Inertial waves} 
\label{secInertial}


In the absence of a warp, the linearised governing equations allow for solutions with either an inertial or an acoustic character. In particular, it is possible to decompose perturbations to the basic state into Fourier-Hermite modes:
\begin{eqnarray}
\delta v_x(x,z,t) = \sum_{m=-\infty}^\infty \sum_{n=0}^\infty u_{m,n}(\tau) \mathrm{He}_n(z)\exp\left(\frac{2\pi im x}{L_x}\right),\\
\delta v_y(x,z,t) = \sum_{m=-\infty}^\infty \sum_{n=0}^\infty v_{m,n}(\tau) \mathrm{He}_n(z)\exp\left(\frac{2\pi im x}{L_x}\right),\\
\delta v_z(x,z,t) = \sum_{m=-\infty}^\infty \sum_{n=1}^\infty w_{m,n}(\tau) \mathrm{He}_{n-1}(z)\exp\left(\frac{2\pi im x}{L_x}\right),\\
\delta h(x,z,t) = \sum_{m=-\infty}^\infty \sum_{n=0}^\infty h_{m,n}(\tau) \mathrm{He}_n(z)\exp\left(\frac{2\pi im x}{L_x}\right),
\end{eqnarray}
where $\mathrm{He}_n$ is the $n$th Hermite polynomial and $L_x$ is the horizontal size of the domain. For a single Fourier-Hermite mode $(m, n)$ the solution $\propto \exp(i\omega \tau)$ where $\omega$ is given, in the inviscid case, by 
\begin{equation}
(-\omega^2+n)\left[-\omega^2 + 2(2-q)\right] - k_x^2 \omega^2=0,
\label{eq:dispersion}
\end{equation} 
where $k_x = 2\pi m \cs/(\Omega L_x)$ is a dimensionless wave number. The low frequency branch of (\ref{eq:dispersion}) is the branch of inertial waves, and in the absence of a warp the various modes are not coupled. The same is true in the presence of viscosity as long as the bulk viscosity $\alpha_b = 2\alpha/3$. It has been found \citep{OL13b} that two inertial waves can couple to the laminar flow induced by the warp, resulting in exponential growth. It is therefore important that inertial waves are captured accurately by the numerical scheme.   

In order to test the ability of the method to sustain inertial waves, we set up an unwarped Keplerian box ($\warp=0$, $q=1.5$), and set up a linear inertial wave (amplitude $10^{-5} \cs$) with $k_x=1.6$ and $n=1$. The horizontal domain is chosen such that two of these waves fit in the domain: $L_x = 4\pi \cs/(\Omega k_x) \approx 7.854\, \cs/\Omega$. The results for $\alpha=0$ are shown for various resolutions in figure \ref{fig:FH_inertial}. The temporal frequency of the inertial wave is $\approx 0.48$, so that we follow the wave for $\approx 40$ periods. During this time, at lowest resolution the wave amplitude is reduced by $50\%$. The major source of error appears to be the dimensionally split nature of the code, in combination with a flow pattern that does not follow the numerical grid \citep[see e.g.][]{astrix}. High resolution is necessary in order to follow the growth of inertial waves, in particular for small growth rates. When viscosity is included, the waves are damped on a time scale $\cs^2\Omega \tau_d \sim k_x^2 /\alpha$. In this case, numerical convergence is easier to obtain, and for $\alpha=0.01$ results with $N=128$ are indistinguishable from those obtained with $N = 256$.   
 

\section{Results}
\label{secRes}


\subsection{Non-Keplerian disc ($q=1.6$)}

While the main focus of this paper is on Keplerian discs with $q=1.5$, we start off by briefly showing results for a disc with $q=1.6$. The non-Keplerian case is slightly easier to set up, as it requires no viscosity and the amplitudes of the laminar motions are much lower for a given warp amplitude compared to the Keplerian case. Initial conditions are given by the laminar warp motion, on top of which we introduce two $k_x=1.6$ inertial waves with $n=1$ and $n=2$ of velocity amplitude $10^{-5}\cs$. The evolution of the $n=1$ Fourier-Hermite component of the vertical velocity is shown in figure \ref{fig:FH_sh16} for $\warp=0.01$ (top panel) and $\warp=0.04$ (bottom panel) for three different resolutions, together with the expected growth rate obtained from linear theory. For $\warp=0.01$, the growth rate is relatively small which means that quite high resolution is needed to match the expected growth rate. This is because the instability is battling against numerical diffusion of the inertial waves. For $N=256$ and $N=512$ the linear growth rate is recovered nicely. The case with $\warp=0.04$ has a much faster linear growth rate, which reduces the resolution requirements to an extent that the linear growth rate can now be recovered for $N=128$.

The torque components evolve erratically with time, as illustrated in figure \ref{fig:Q_sh16}. As expected from the laminar flows, initially only the $Q_3$ component is non-zero. In the end, all three warp amplitudes considered settle on an average value of $Q_3$ that is slightly reduced compared to the initial value, but with a large spread. Interestingly, also the $Q_2$ component, governing roughly speaking the diffusion of the warp, reaches similar values on average for all three warp amplitudes. Only for $Q_1$ do we observe a clear trend with $\warp$, with $Q_1$ increasing with increasing $\warp$. While there are large short time scale oscillations visible, on average $Q_1$ settles to a positive value. A positive value of $Q_1$ indicates a reversal of the angular momentum flux from a usual accretion disc, which means one could expect the mass distribution to behave in an antidiffusive manner, leading to a breakup of the disc into disjoint rings. However, the effect is very weak in this case.   

The absence of viscosity in principle allows for structure on very small scales. In figure \ref{fig:velx_sh16} we show snapshots of $\sqrt{\rho} |{\bf \delta v}|$, where ${\bf \delta v}$ is the velocity perturbation on top of the laminar motion, which is a measure of wave energy.  The wave energy increases with warp amplitude, and there is a tendency for more small-scale structure at larger $\warp$. For $\warp=0.01$, the final state is dominated by the $k_x=1.6$ mode, independent of whether we initialize a single $k_x=1.6$ mode or start from white noise. While smaller scale inertial waves (larger wave numbers) have higher growth rates, these tend to saturate at low amplitudes, leaving enough of the laminar motions for the smaller wave number modes to feed off and grow. As a result, the nonlinear state is well captured (as indicated by the torque components) even at modest resolution of $N=128$, despite the fact that this resolution was unable to accurately reproduce the linear growth rates (see figure \ref{fig:FH_sh16}). 

As the amplitudes of the laminar flow remain relatively modest for $q=1.6$, larger warp amplitudes can be considered without difficulty. In fact, the main limiting factor, in particular in runs without any explicit viscosity, is the small time steps necessary because of large velocities in the upper layers of the box. Here we note in particular that while no laminar solutions can be found for $\warp>0.261$ due to a nonlinear resonance \citep{OL13a}, with regards to the nonlinear state this value of $\warp$ is not special. There exists a nonlinear state for $\warp=0.27$, which can be reached either by starting without any laminar flow or by starting from the nonlinear state at $\warp=0.26$ and increase the warp amplitude. The final state at $\warp=0.27$ is very similar to that at $\warp=0.26$ in terms of torque components and velocity structure.


\subsection{Keplerian disc ($q=1.5$)}

\begin{figure}
\includegraphics[width=\columnwidth]{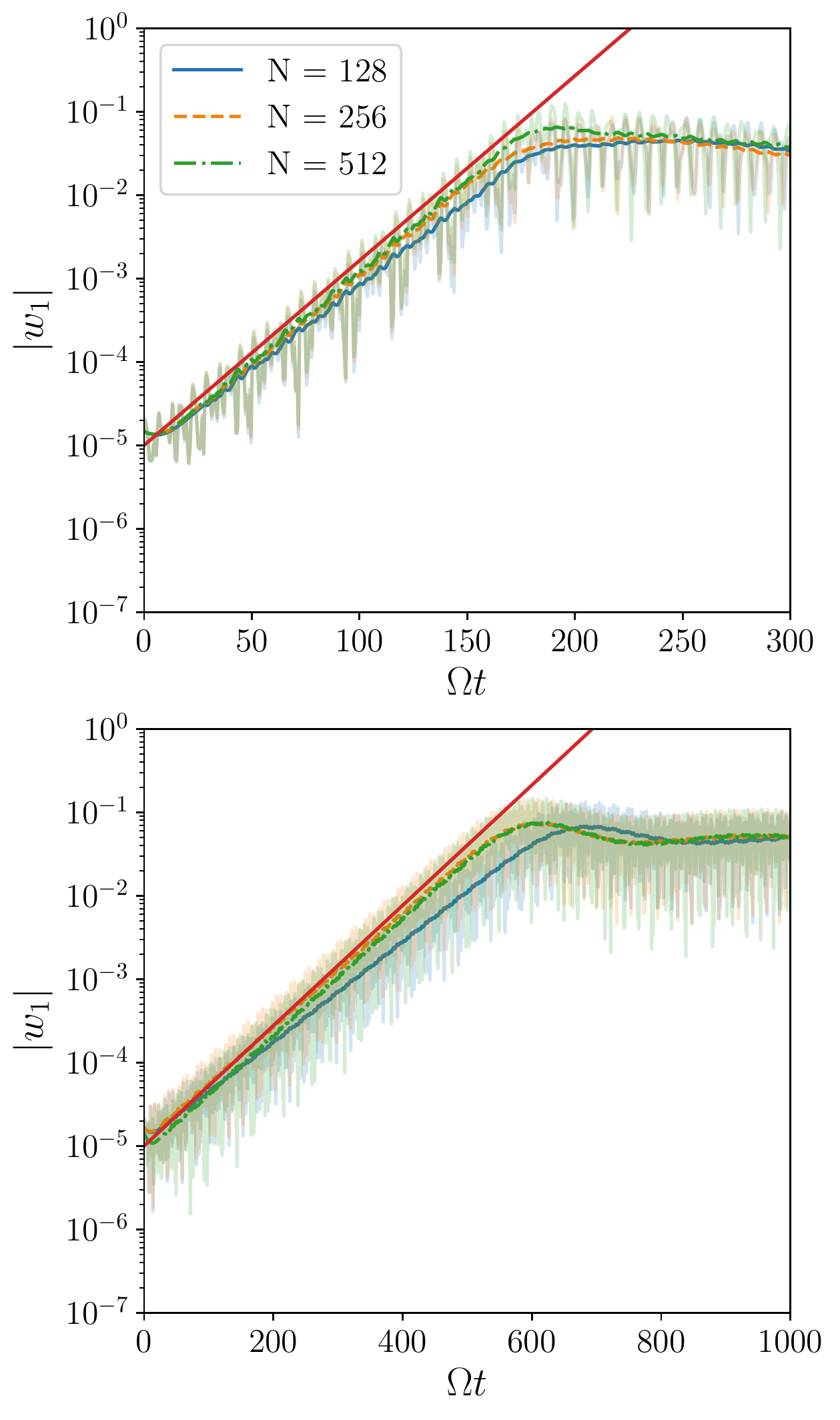}
\caption{Evolution of the magnitude of the $n=1$, $k_x=1.6$ Fourier-Hermite component of the vertical velocity in a Keplerian $q=1.5$ disc with $\warp=0.001, \alpha=0.001$ (top panel) and $\warp=0.01, \alpha=0.01$ (bottom panel) for different resolutions. The solid red line indicates the growth rate determined from a linear calculation.}
\label{fig:FH_sh15}
\end{figure}

\begin{figure}
\includegraphics[width=\columnwidth]{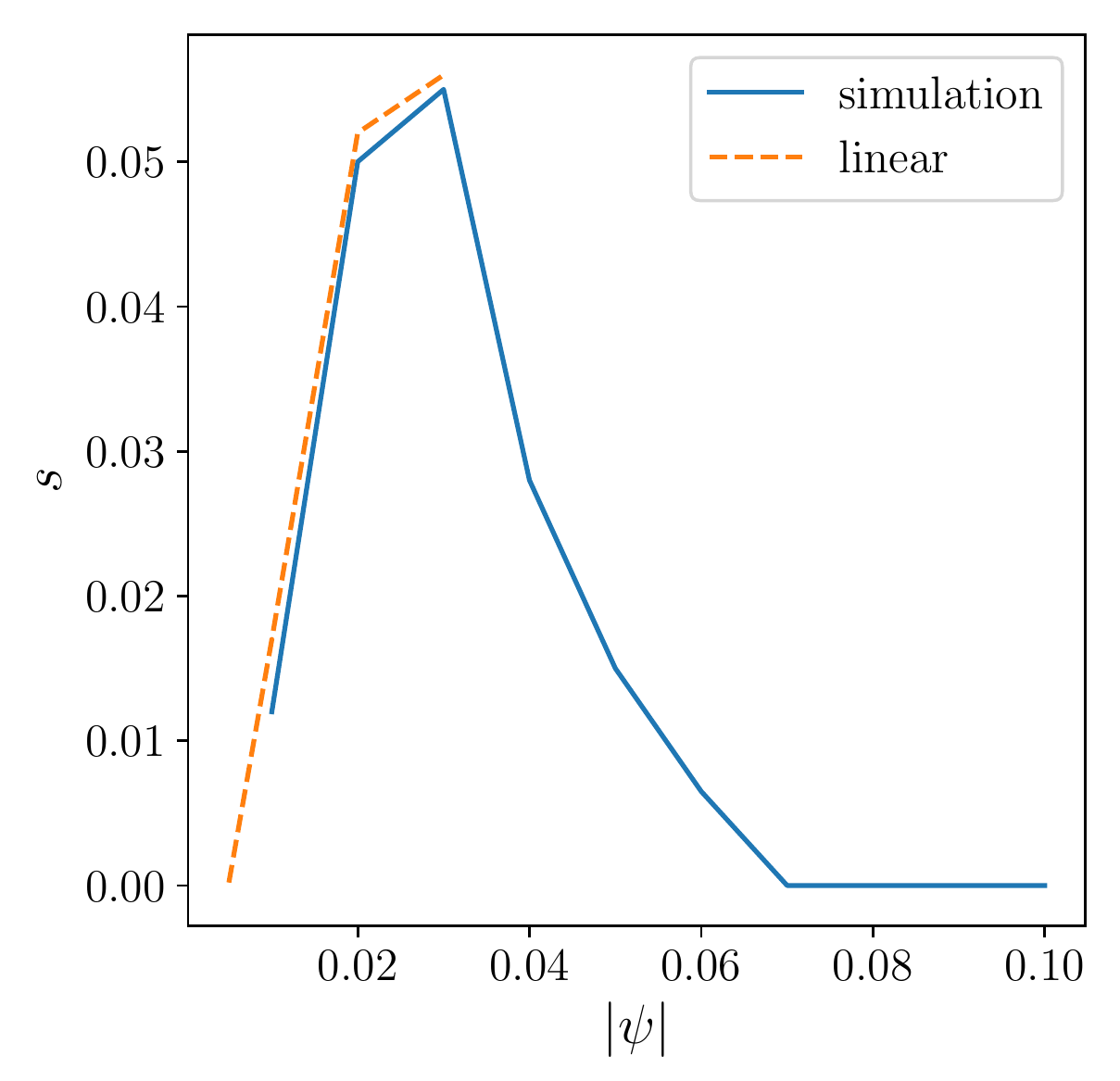}
\caption{Growth rate, maximised over $k_x$, as a function of warp amplitude for $q=1.5$ and $\alpha=0.01$. Results obtained by hydrodynamical simulations as well as linear calculations are shown. Note that beyond $\warp=0.03$, no reliable linear growth rates were found by the latter method.}
\label{fig:growth_nu-2}
\end{figure}

\begin{figure}
\includegraphics[width=\columnwidth]{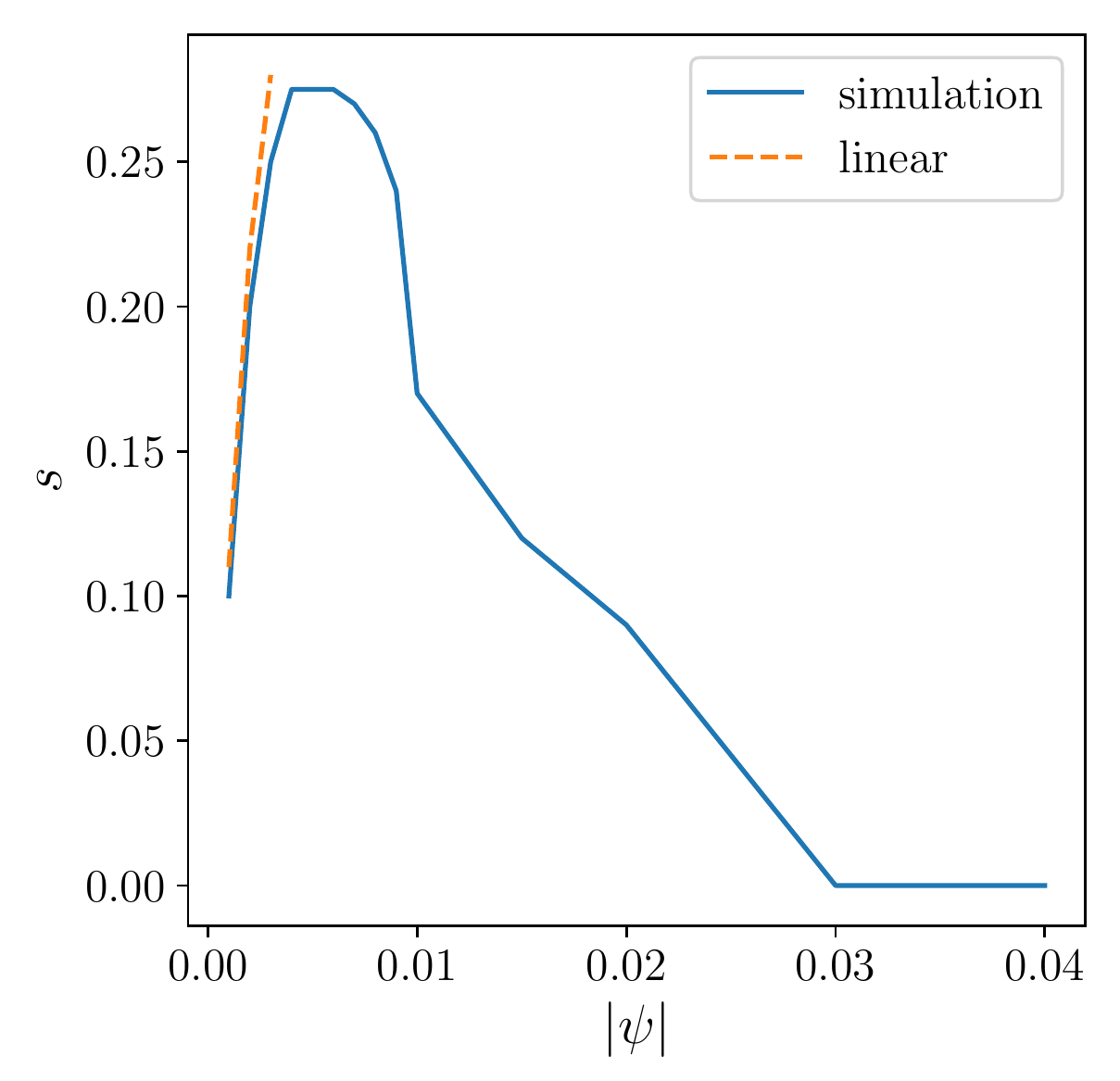}
\caption{Growth rate, maximised over $k_x$, as a function of warp amplitude for $q=1.5$ and $\alpha=0.001$. Results obtained by hydrodynamical simulations as well as linear calculations are shown. Note that beyond $\warp=0.003$, no reliable linear growth rates were found by the latter method.}
\label{fig:growth_nu-3}
\end{figure}

\begin{figure}
\includegraphics[width=\columnwidth]{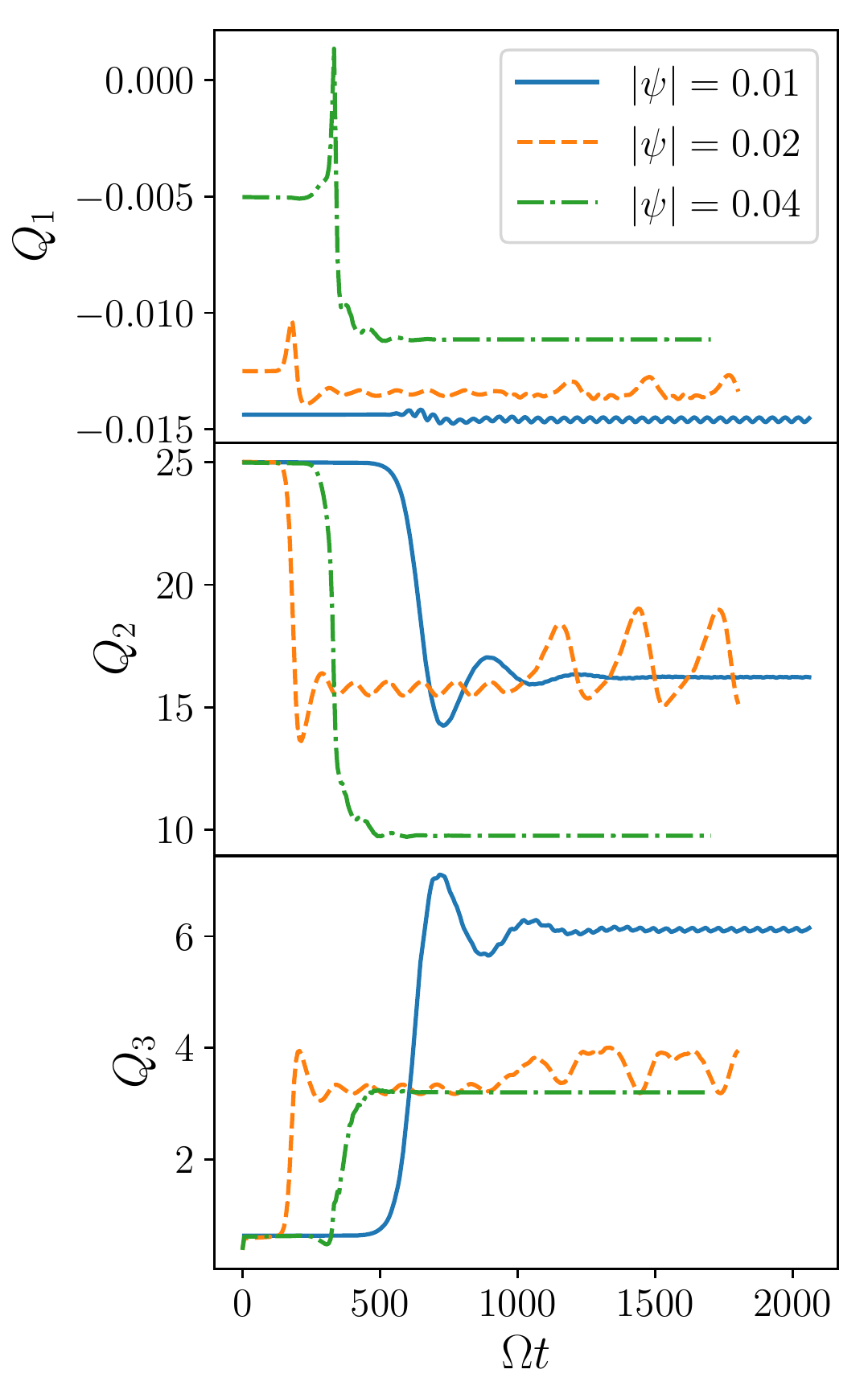}
\caption{Evolution of the torque components for a disc with $q=1.5$ and $\alpha=0.01$ for three different warp parameters, all at resolution $N=512$.}
\label{fig:Q_nu-2}
\end{figure}

\begin{figure*}
\includegraphics[width=\textwidth]{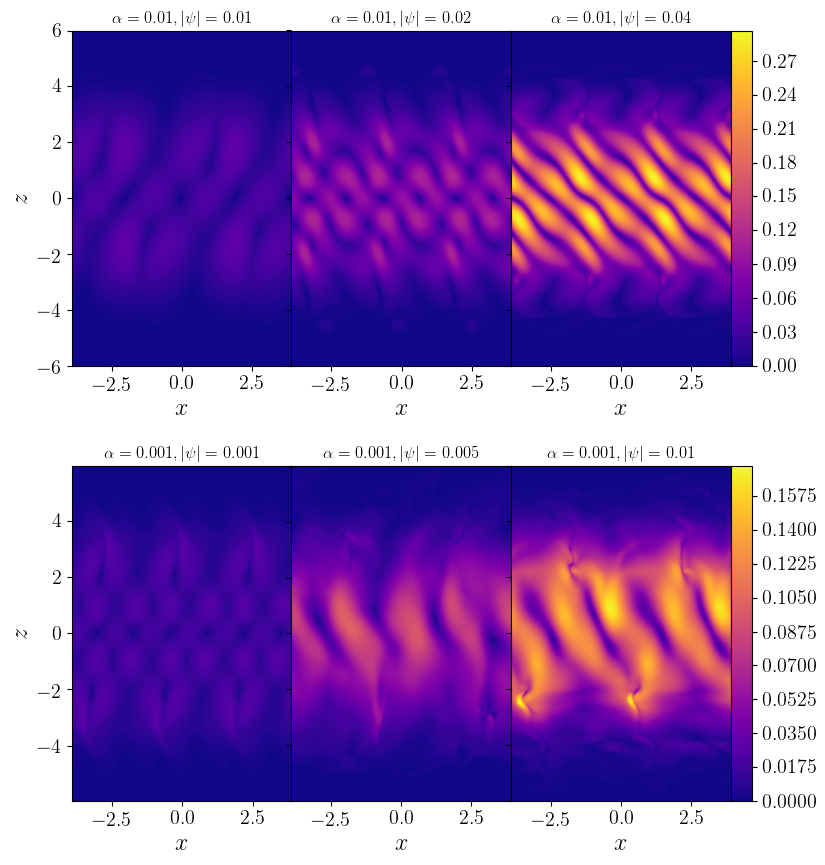}
\caption{Snapshots of $\sqrt{\rho}|{\bf \delta v}|$, which is a measure of wave energy, for a disc with $q=1.5$ and $\alpha=0.01$ (top row) and $\alpha=0.001$ (bottom row) in the nonlinear phase, for three different warp amplitudes, all at resolution $N=512$. Top left panel: $\warp=0.01$ at $\Omega t=600\pi$. Top middle panel: $\warp=0.02$ at $\Omega t = 200\pi$. Top right panel: $\warp=0.04$ at $\Omega t = 600\pi$. Bottom left panel: $\warp=0.001$. Bottom middle panel: $\warp=0.005$. Bottom right panel: $\warp=0.01$. All three bottom panels show results at $\Omega t = 400\pi$.}
\label{fig:velx_visc}
\end{figure*}

\begin{figure}
\includegraphics[width=\columnwidth]{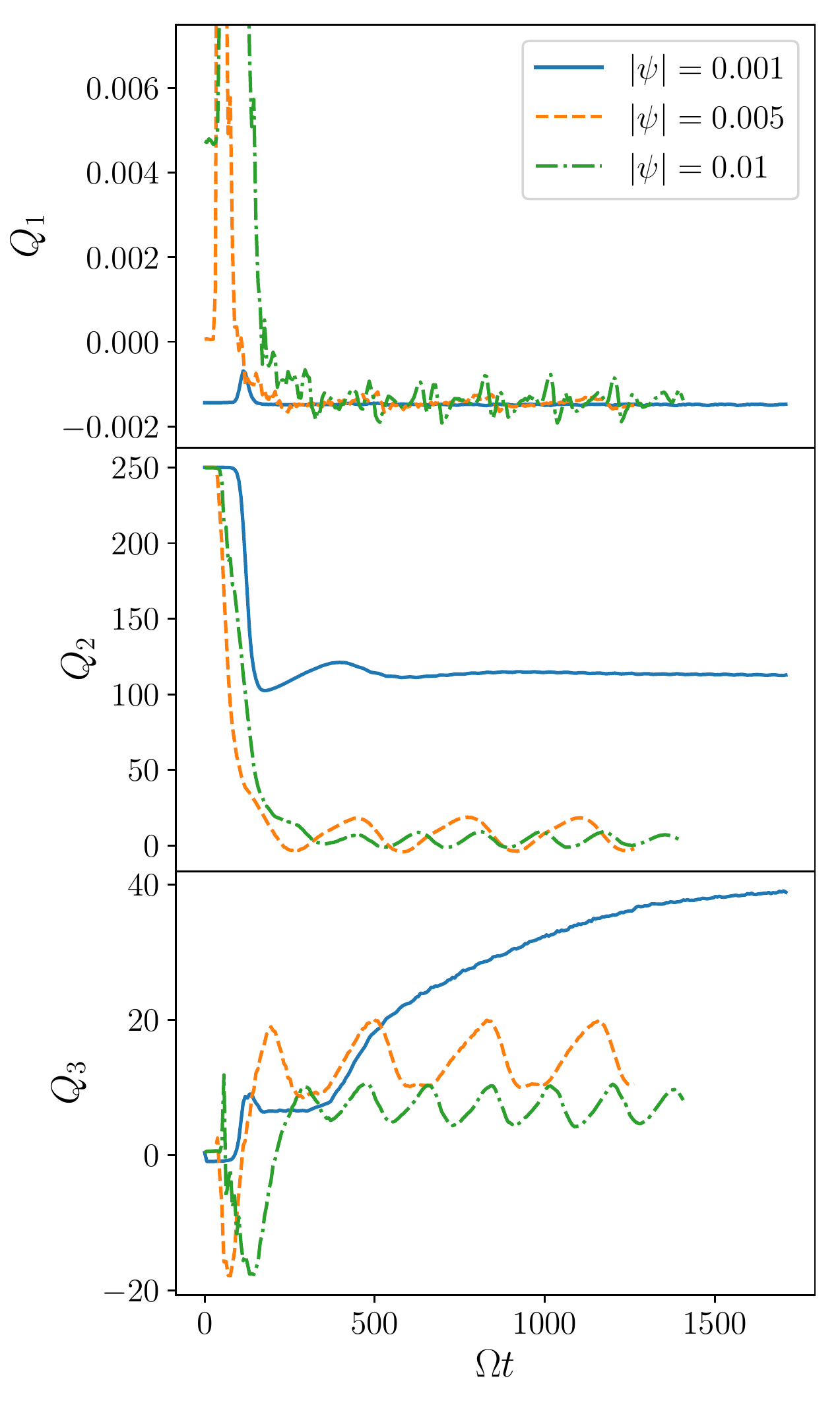}
\caption{Evolution of the torque components for a disc with $q=1.5$ and $\alpha=0.001$ for three different warp parameters, all at resolution $N=512$.}
\label{fig:Q_nu-3}
\end{figure}

\begin{figure}
\includegraphics[width=\columnwidth]{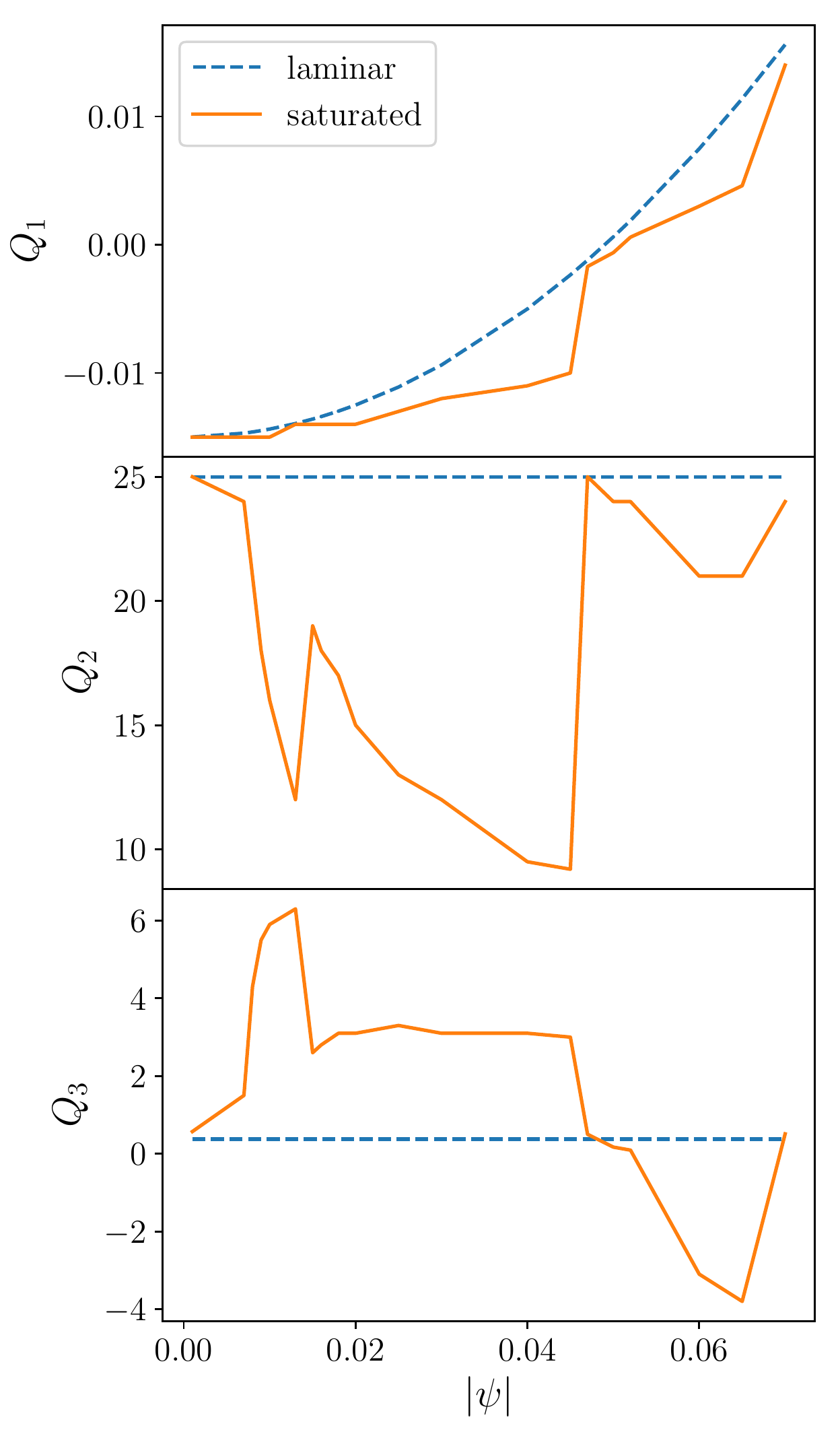}
\caption{Variation of torque components with warp amplitude for $q=1.5$ and $\alpha=0.01$, for both the initial laminar state and the saturated nonlinear state.}
\label{fig:Qpsi_nu-2}
\end{figure}

\begin{figure}
\includegraphics[width=\columnwidth]{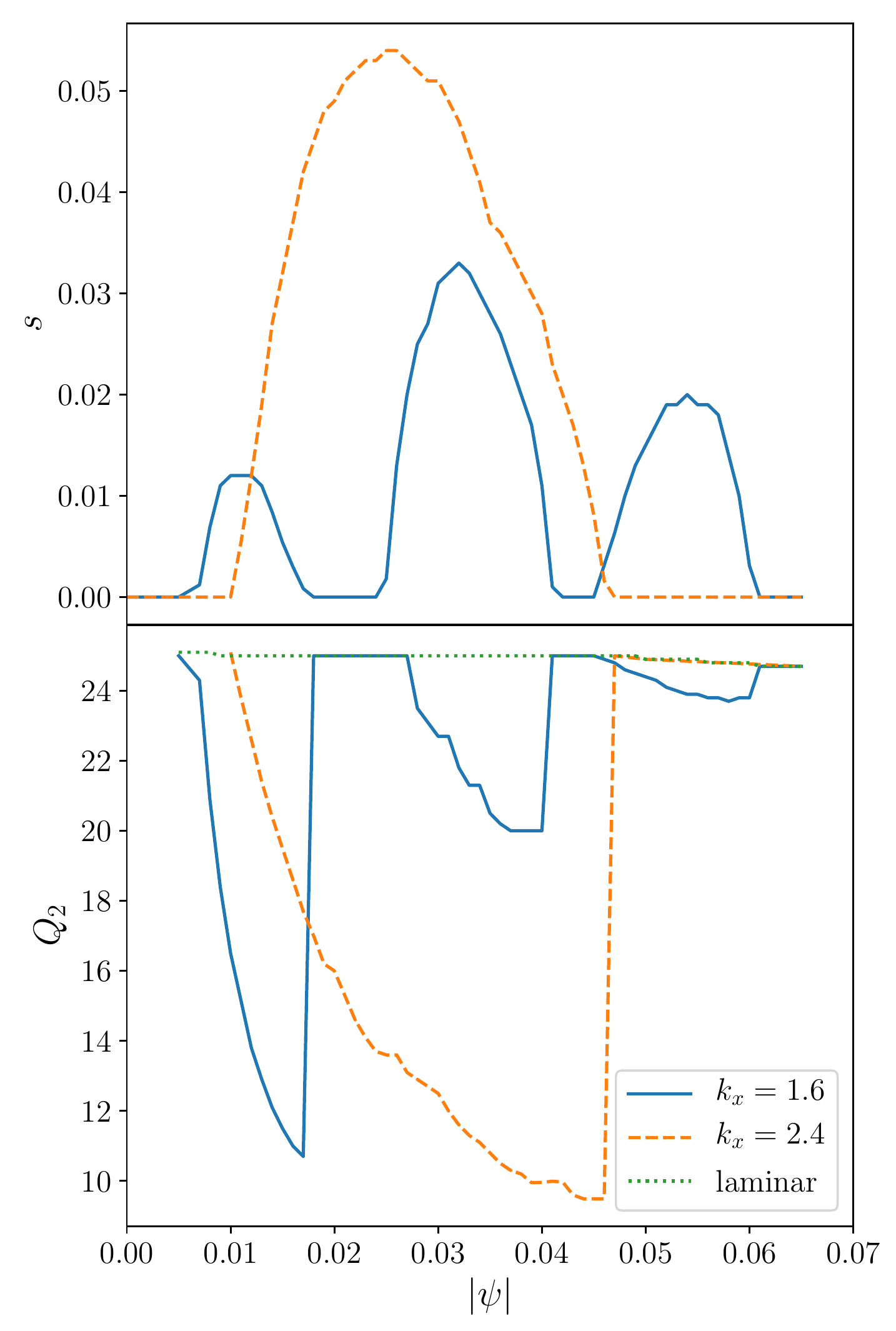}
\caption{Growth rates (top panel) and torque component $Q_2$ (bottom panel) for $q=1.5$, $\alpha=0.01$ and initial conditions selecting a single value of $k_x$.}
\label{fig:growth_splitk}
\end{figure}

\begin{figure}
\includegraphics[width=\columnwidth]{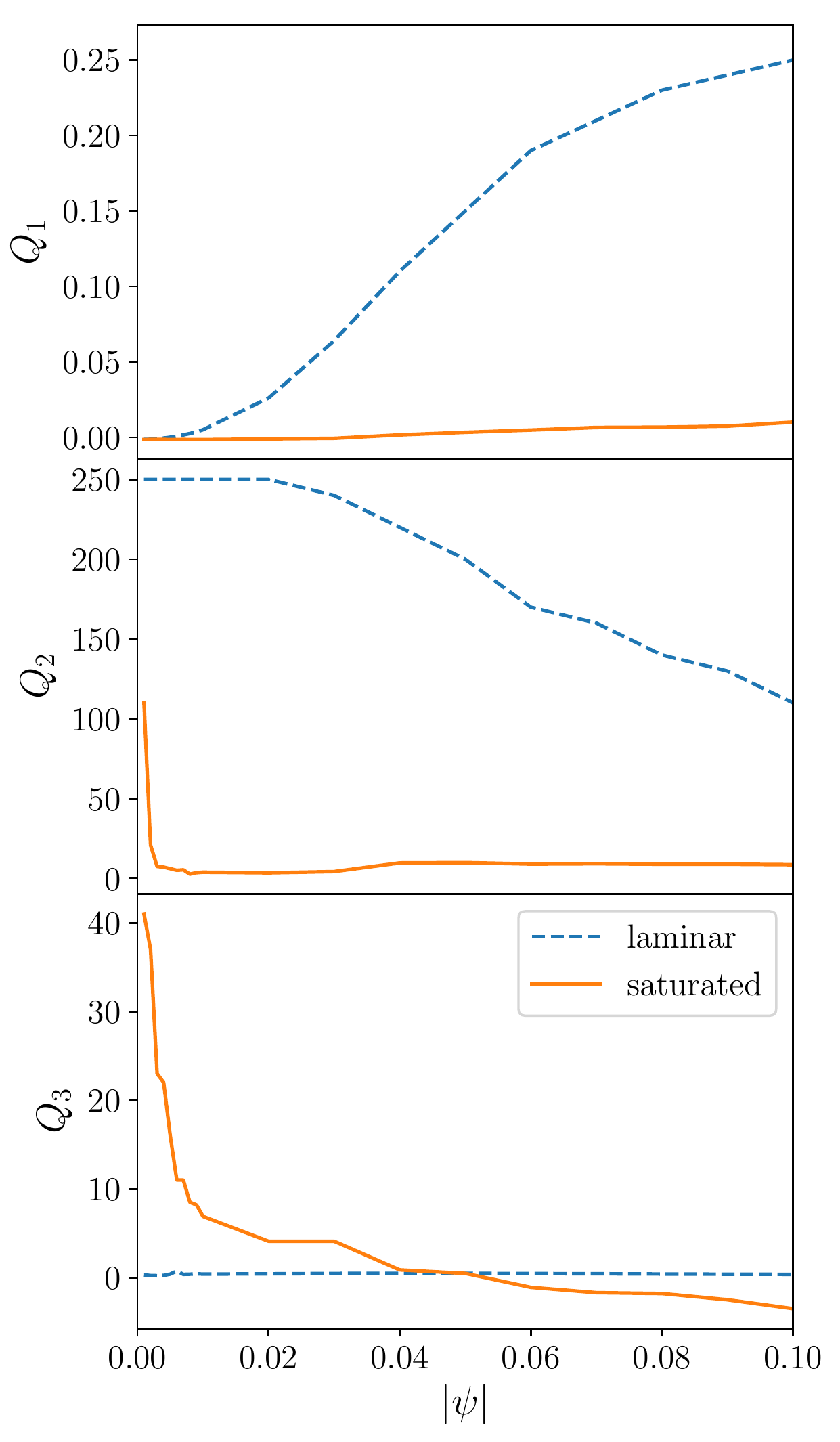}
\caption{Variation of torque components with warp amplitude for $q=1.5$ and $\alpha=0.001$, for both the initial laminar state as the saturated nonlinear state.}
\label{fig:Qpsi_nu-3}
\end{figure}

We now turn our attention to the interesting Keplerian case that has $q=1.5$. For $q=1.5$, viscosity is needed in order for laminar flow solutions to exist \citep{OL13a}. We consider two viscosity coefficients: $\alpha = 0.001$ and $\alpha=0.01$, always leaving the bulk viscosity zero. 


\subsubsection{Growth rates}


In figure \ref{fig:FH_sh15} we show the evolution of the $k_x=1.6$, $n=1$ Fourier-Hermite component of the vertical velocity for $\warp=\alpha=0.001$ (top panel) and $\warp=\alpha=0.01$ (bottom panel), for three different resolutions together with the growth rate obtained from a linear calculation. The linear growth rate is much larger in the case of $\alpha=0.001$, which makes it easier to capture accurately at low resolution, so that even $N=128$ does a reasonable job of reproducing the linear growth rate. The slower growth rate when $\alpha=0.01$ is only captured accurately for $N=256$ and above, even though the saturated state is very similar in all cases.

For $\alpha >0$, growth rates are reduced for large wave numbers, and for a given warp amplitude $\warp$ and viscosity $\alpha$, we can calculate the growth rate maximized over $k_x$ either by a linear calculation \citep{OL13b} or through direct numerical simulation by initializing the velocities with white noise and measuring the growth of the perturbed energy. Note that in the latter case, we obtain the maximum over all $k_x$ that fit into the box. In our standard setup, the minimum $k_x$ we can consider is $k_x=0.8$ (one wave fits in the box), and the allowed wave numbers are therefore $k_x = 0.8, 1.6, 2.4, 3.2$, etc. While in the picture of the pure parametric instability this may pose a problem, as a particular value of $k_x$ is required for the waves to couple \citep{gammie00, OL13b} which may not fit into the box, in practice, the instability bands for the values of $\alpha$ considered here are wide enough so that our sampling gives a good estimate of the maximum growth rate.   

An issue with solving the linearized equations for the Fourier-Hermite components that was flagged up in \cite{OL13b} is that for strong laminar flows the spectrum fails to converge and no reliable growth rates can be obtained. The difficulty may lie in the Hermite basis being unable to handle warped discs with strong laminar flows, but it may also be that the oscillatory shear of the laminar flow may prevent the formation of axisymmetric eigenmodes altogether. An independent implementation of the numerical method based on Floquet theory discussed in \cite{OL13b} shows the same behaviour: for $\alpha=0.01$ we can find no reliable growth rates for $\warp > 0.03$, while for $\alpha=0.001$ we can find no reliable growth rates for $\warp > 0.003$. 

The linear results, together with growth rates measured from the hydrodynamic simulations, are displayed in figure \ref{fig:growth_nu-2} for $\alpha=0.01$. For $\warp \leq 0.03$, the results obtained with the two methods agree very well. Interestingly, we can numerically find positive growth rates where the linear solver fails to converge. This could indicate that at least part of the problem with the linear solver lies indeed with the use of the Hermite basis. However, the measured growth rates quickly decline towards larger values of $\warp$ and in the end for $\warp \geq 0.07$ no growing modes are found at all. While it is possible that the hydrodynamic simulations suppress very small growth rates due to numerical damping, it is worth noting that the growth rate at $\warp = 0.04$ is already reasonably well captured with our lowest resolution $N=128$. In addition, for this relatively large value of $\alpha$ one expects large wave numbers to be strongly damped, so that one would not expect the necessary resolution to be extremely large. The main requirement is for the laminar flows to be accurately represented, for which one needs $N = 256$ at least for the larger warp amplitudes. Higher resolution and larger horizontal box sizes did not yield any growth rates for $\warp \geq 0.07$, which supports the conclusion that the shear imposed by the laminar flows prevents the formation of axisymmetric eigenmodes. 

Lowering the viscosity to $\alpha=0.001$ yields a similar picture, although because of the stronger laminar flows difficulties arise at smaller warp amplitudes. In figure \ref{fig:growth_nu-3} we show again growth rates obtained from both the linearized equations and hydrodynamic simulations. No reliable growth rates were found from the linearized equations for $\warp > 0.003$, while the hydrodynamic simulations show growing modes up to $\warp=0.03$. Beyond $\warp=0.03$, no growing modes were found (but see section \ref{sec:hyst} where we find a \emph{nonlinear} instability), again despite increasing the resolution up to $N=2048$ and changing the horizontal size of the box. Note that, unlike in the case with $\alpha=0.01$, growth rates remain high after the linear results break down, before decreasing for $\warp > 0.01$.  


\subsubsection{Saturated state}


Once the perturbations grow to significant amplitude, the torque components $Q_1$, $Q_2$ and $Q_3$ start to be affected. This is because the perturbations feed off the laminar flow, which can therefore be strongly reduced in the saturated state. The torque components in the saturated state are important indicators of subsequent evolution of the warp. In particular, when $Q_2>0$ it can be interpreted as a diffusion coefficient for the warp. From equation (\ref{eqQ2approx}), it is clear that laminar flows at small viscosity would lead to very fast evolution of the warp because of the factor $\alpha$ in the denominator. However, as these laminar flows were found to be unstable, one should look at the saturated state.

In figure \ref{fig:Q_nu-2} we show the evolution of the three torque components for the high viscosity case $\alpha =0.01$ for three different warp amplitudes. At early times, the laminar flows dominate the torques, and excellent agreement is found with the results of \cite{OL13a}. The onset of the nonlinear phase is marked most clearly by a sharp drop in $Q_2$, accompanied by a drop in $Q_1$, which is barely visible for $\warp=0.01$, and an increase in $Q_3$. Towards late times, the disc settles into a saturated state, usually with well-defined values for the torque components (the exception being $\warp=0.02$, which we will comment on below). The drop in $Q_1$ signals that inward angular momentum transport is enhanced, up to a factor of 2 in the case of $\warp=0.04$, while for smaller warps the effect is reduced. The drop in $Q_2$ means that the diffusion of the warp is reduced by roughly a factor 2 in all cases. The third torque component $Q_3$ increases in all cases. While both the reduction in $Q_2$ and the increase in $Q_3$ are modest, since they are of comparable magnitude in the saturated state one might expect the warp to behave in a more wave-like manner compared to the laminar state, which has $Q_2 > Q_3$. We will see below that this effect is more pronounced in the case where the viscosity is smaller.

The saturated state is usually dominated by modes of a single value of $k_x$. This is illustrated in the top panels of figure \ref{fig:velx_visc}, where we show $\sqrt{\rho} |{\bf \delta v}|$, where ${\bf \delta v}$ is the velocity perturbation on top of any possible (horizontally uniform) laminar flow. This quantity is a measure of wave energy. The smallest warp amplitude $\warp= 0.01$ is seen to be dominated by $k_x=1.6$ (note that because of the absolute value four maxima, as observed in the top left panel of figure \ref{fig:velx_visc}, mean two wavelengths over the length of the box, which translates to $k_x=1.6$), while the two larger warp amplitudes are dominated by modes with $k_x=2.4$. In all cases, density perturbations remain below $\sim 10\%$ in the mid plane. Note that the wave energy goes to zero rapidly for $|z| > 4 c_s/\Omega$, far enough away from the vertical boundaries at $|z|= 6c_s/\Omega$ to be confident that the solution is unaffected by the boundaries. 

The torque components for $\warp=0.02$ show a second transition around $\Omega t = 1000$, while all others settle into a steady saturated state after the onset of nonlinearity with no apparent further changes. In the case of $\warp=0.02$, before this second transition, the saturated state is dominated by modes with $k_x=2.4$, as shown in figure \ref{fig:velx_visc}. However, in this state modes with $k_x=1.6$ can still couple to the reduced laminar flow and grow, albeit more slowly than they would have in the presence of the full laminar flow. Once these new modes reach nonlinear amplitudes, the second transition happens where the modes of different $k_x$ compete for dominance. No such growth of additional modes was observed for $\warp = 0.01$ or $\warp=0.04$.   

For the low viscosity case $\alpha=0.001$, we find similar but more extreme results. In figure \ref{fig:Q_nu-3} we show the evolution of the torque components for three different values of $\warp$. Since for the laminar flows, $Q_2 \propto 1/\alpha$, initially $Q_2$ is a factor of 10 larger than for $\alpha=0.01$. In the saturated state however, even a modest warp of $\warp=0.005$ can significantly reduce the amplitude of the laminar flow, much more so than in the more viscous case of $\alpha=0.01$. Both $\warp=0.005$ and $\warp=0.01$ show a reduction in $Q_2$ of roughly a factor $100$, which should significantly reduce the diffusion of these warps. 

The laminar flows associated with the warps shown in figure \ref{fig:Q_nu-3} predict a reversal of angular momentum flux for $\warp=0.01$ so that $Q_1>0$ initially ($\warp=0.005$ being a borderline case). Such a situation would see the mass evolve in an anti-diffusive manner, probably leading to a breakup of the disc into distinct rings, as observed in for example \cite{nixon12}. In the saturated state, however, all three warp amplitudes have $Q_1 < 0$, where also for $\warp=0.001$ the saturated value of $Q_1$ is more negative than the initial, laminar, value. The third torque component $Q_3$ increases in magnitude roughly by an order of magnitude. This means that while in the laminar state we have $|Q_2| \gg |Q_3|$, in the saturated state we have $|Q_2| < |Q_3|$, and one might expect a more wave-like behaviour of the warp.
 
Similar to the higher viscosity case, the saturated state is dominated by either $k_x=1.6$ or $k_x=2.4$ (see bottom panels of figure \ref{fig:velx_visc}). The largest warp amplitude ($\warp=0.01$) shows some small-scale structure around $z \approx \pm 2 c_s/\Omega$. The mid plane perturbations in density remain modest, up to $\sim 5\%$ for $\warp=0.01$. Similar to the higher viscosity case, the wave energy is essentially concentrated in $|z|<4 c_s/\Omega$.
 

\subsubsection{Hysteresis}
\label{sec:hyst}


We are now in a position to survey the parameter space, finding $Q_i = Q_i(\warp,\alpha, q)$, where $Q_i$ stands for any of the three torque components, in the saturated state. We will only consider the Keplerian case, $q=1.5$, and consider two values of $\alpha=0.01$ and $\alpha=0.001$ as before. 

In figure \ref{fig:Qpsi_nu-2} we show the three torque components as a function of warp amplitude for the case of $\alpha=0.01$. The laminar results show the expected behaviour of $Q_1$ increasing quadratically with $\warp$, while both $Q_2$ and $Q_3$ are nearly constant \citep{OL13a}.  For $\warp > 0.05$ we have that $Q_1>0$ and therefore angular momentum flux reversal for the laminar solution. In the saturated state, $Q_1$ is always below the laminar value, but  all the same turns positive for $\warp > 0.05$. Moreover, $Q_1$ never goes below $-3\alpha/2$, which would be the value for an unwarped laminar disc. Similarly, the value of $Q_2$ in the saturated state is always below the laminar value, indicating that the warp will diffuse more slowly than would be the case for purely laminar flow. The minimum value of $Q_2=9.2$ is reached for $\warp=0.045$. 

Several transitions can be identified in the saturated curves in figure \ref{fig:Qpsi_nu-2}, most notably at $\warp=0.008$, where there is a discontinuity in $d Q_2/d\warp$, at $\warp=0.015$ where there is a jump in $Q_2$ and finally at $\warp=0.05$ where $Q_2$ jumps almost to its laminar value. These transitions are associated with a change in the dominant wave number. For $\warp<0.015$ the saturated state is dominated by $k_x=1.6$, switching to $k_x=2.4$ for $0.015<\warp < 0.05$. While it looks like at $\warp=0.05$ there are no growing modes and the laminar state is recovered, there are growing $k_x=1.6$ modes but they saturate at very low amplitude. Beyond $\warp=0.05$, the solution is dominated by modes with $k_x=0.8$. The growth rate of these modes is relatively slow and they may involve higher-order mode couplings \citep{OL13b}.

\begin{figure*}
\includegraphics[width=\textwidth]{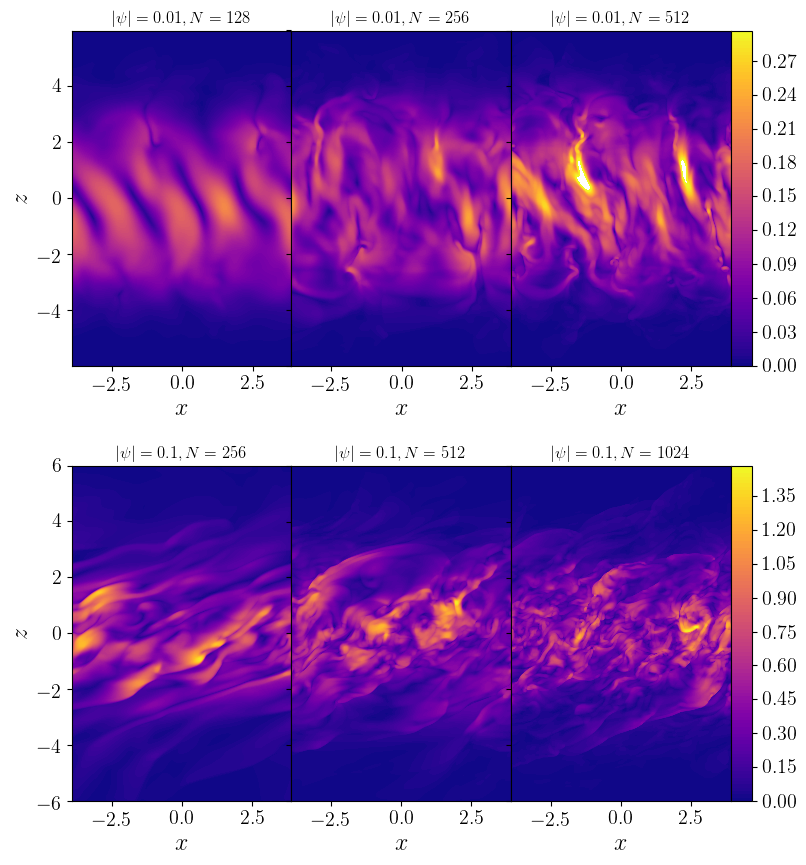}
\caption{Snapshots of $\sqrt{\rho}|{\bf \delta v}|$, which is a measure of wave energy, for an inviscid disc with $q=1.5$ in the nonlinear phase for two different warp amplitudes, each at three different resolutions, all at  $\Omega t=200\pi$.}
\label{fig:velx_nolaminar}
\end{figure*}

The low level of saturation around $\warp=0.05$ does not correlate with a reduced growth rate. As can be seen in figure \ref{fig:growth_nu-2}, the growth rate is above $0.01$ for this warp amplitude. The low level of saturation can be understood from the growth rates, however, if we look for specific values of $k_x$, as shown in figure \ref{fig:growth_splitk}. Modes with $k_x=1.6$ can grow in three distinct regions of $\warp$. The amplitude at which the modes saturate depends on a competition between growth from feeding off the laminar flows and viscous or nonlinear damping. A mode with $k_x=1.6$ growing in a $\warp=0.015$ warp can grow until it becomes nonlinear, at which point the laminar flows are significantly reduced, as indicated by the reduction in $Q_2$. The reduced laminar flows correspond to a lower value of $\warp$, but from figure \ref{fig:growth_splitk} we see that to the left of $\warp=0.015$ the growth rate is actually larger, making it possible for the mode to grow to larger amplitude. If the laminar flows are reduced too much, however, the growth rate first starts to decline before growth shuts off completely. This puts a constraint on the maximum amplitude these modes can reach. In the case of $k_x=1.6$ and $\warp=0.015$, the laminar flows can be reduced until they correspond to $\warp=0.005$, which is a significant reduction leading to large saturated wave amplitude, as indicated by the low saturated value of $Q_2\approx 11$. In contrast, starting at $\warp=0.06$, the laminar flows can only be reduced to $\warp\approx 0.05$, which is a small relative reduction leading to a low saturated wave amplitude, as indicated by a saturated value of $Q_2 \approx 23$ that is close to the laminar value. Note that because there is a large single region where modes with $k_x=2.4$ can grow, these modes always saturate at large amplitudes.

Such distinct regions of non-zero growth rates as seen in figure \ref{fig:growth_splitk} are not expected from the simple three-wave coupling model \citep{gammie00, OL13b}, which is valid for $\warp$ small enough so that the shear from the laminar flow is much less than the Keplerian shear. The linear calculations presented in \cite{OL13b} show that for larger warp amplitudes, the instability bands can shift to different values of $k_x$(see for example their figure 5, where $k_x=1.6$ is inside a band of instability for $\warp=0.01$, but not for $\warp=0.02$, which is consistent with our results). We find that at even larger warp amplitudes, instability can return to $k_x=1.6$, but in the end disappears completely. A shift in frequency of the most unstable modes towards larger warp amplitudes can also be obtained from the incompressible model presented in \cite{gammie00}, but for this model the instability band widens at the same time so that $k_x=1.6$ always remains unstable. It is therefore likely that compressibility plays an important role in this phenomenon.

Lowering the viscosity to $\alpha=0.001$ reveals new interesting behaviour. While we could not find any growing modes for $\warp > 0.03$ (see figure \ref{fig:growth_nu-3}), it is possible to find a nonlinear state for warp amplitudes up to at least $\warp=0.1$. This state can be reached for example by starting from the saturated state at lower warp amplitude and slowly increasing $\warp$, or by starting without any laminar flow. The system therefore exhibits hysteresis. It takes time to set up the laminar flow either from scratch or from a lower value of $\warp$, and as long as modes can grow to sufficient strength before the full laminar flow is reached, the resulting state will be the saturated nonlinear state rather than the laminar flow. It should be noted that hysteresis appears to be limited to the low viscosity case: for $\alpha=0.01$ we always recover the laminar state for $\warp \geq 0.07$, no matter the initial conditions.

The results for $\alpha=0.001$ are displayed in figure \ref{fig:Qpsi_nu-3}. First thing to note is that since we are considering larger warp amplitudes compared to the more viscous runs with $\alpha=0.01$, the laminar torques start to differ from the quadratic form (for $Q_1$) and the constant value (for $Q_2$) towards larger $\warp$, while $Q_3$ remains roughly constant. The torque component $Q_1$ saturates at a value of $\sim 0.1$, while $Q_2$ starts to decrease for $\warp > 0.01$. As is the case for the more viscous runs, $Q_1 >0$ for larger enough $\warp$, in this case $\warp \geq 0.005$. 

The torque components in the saturated state vary much more smoothly compared to the more viscous runs. This is probably due to more unstable modes being available for $\alpha=0.001$. For all warp amplitudes we find $Q_1$ in the saturated state to be much smaller than in the laminar state. Angular momentum flux reversal occurs for $\warp > 0.04$, and $Q_1$ keeps increasing for larger warp amplitudes. Except for the lowest warp amplitudes $\warp < 0.002$, $Q_2$ is reduced substantially in the saturated state compared to the laminar state. This is most prominent at $\warp \sim 0.01-0.02$, where $Q_2$ is reduced by two orders of magnitude. For larger warp amplitudes the effect is less dramatic, mainly due to the lower values of $Q_2$ in the laminar state. The third torque component shows a large variation with warp amplitude in the saturated state, unlike the laminar state for which $Q_3$ is roughly constant. Note that while $|Q_2| \gg |Q_3|$ in the laminar state for essentially all values of $\warp$, the saturated state has $|Q_2| \lesssim |Q_3|$, perhaps indicating a more wavelike behaviour of the warp.
 
For warp amplitudes $\warp > 0.1$ it becomes more difficult numerically to sustain the nonlinear saturated state. For example, at resolution $N=512$ at $\warp=0.12$, a saturated state can be set up but it ultimately decays towards a laminar flow. The same simulation at a higher resolution $N=1024$ shows no signs of decaying. It may be the case that for $\alpha=0.001$ a saturated state exists for all values of $\warp$ but the regime $\warp>0.2$ will be very expensive to reach computationally.


\subsubsection{Inviscid limit}


Finally, we briefly consider the inviscid limit. In the limit of vanishing viscosity, laminar solutions cease to exist for a Keplerian rotation profile \citep{OL13a}. One may first of all wonder whether a saturated state like those found above still exists. We start from purely Keplerian flow with small amplitude white seed noise, and impose a warp $\warp$ as before. Results after $\Omega t = 200\pi$ are shown in figure \ref{fig:velx_nolaminar}. For both the small ($\warp=0.01$, top row) and large ($\warp=0.1$, bottom row) warp amplitude a saturated state very much like the viscous case is set up. Towards higher resolution, smaller scale flow structures can be seen. These did not feature in the viscous runs, compare for example figure \ref{fig:velx_visc}. However, the measured torque components for $\warp=0.01$ ($Q_1=10^{-5}$, $Q_2=3.0$, $Q_3=6.0$) are independent of resolution and agree very well with the viscous case that has $\alpha=0.001$. The small scale features therefore do not affect the internal torque. The agreement with $\alpha=0.001$ suggests that this viscosity is representative of a ``low viscosity" regime where the torque components do not sensitively depend on the viscosity.    

For the large warp amplitude ($\warp=0.1$, bottom row of figure \ref{fig:velx_nolaminar}), again the measured torque components ($Q_1=0.016$, $Q_2=8.0$, $Q_3=-2.0$) agree well with those obtained for $\alpha=0.001$. However, numerical convergence is more difficult to obtain. While $N=512$ and $N=1024$ give identical results, $N=256$ for example gives $Q_2=13.0$. At even lower resolution $N=128$ (not shown in figure \ref{fig:velx_nolaminar}), the flow settles in an almost laminar state with very low amplitude waves superimposed. This again highlights the strong resolution constraints towards larger warp amplitudes. Even when a completely laminar state does not exist, numerical diffusion will tend to take the flow towards a state where any non-laminar component has very low amplitude.   

\begin{figure}
\includegraphics[width=\columnwidth]{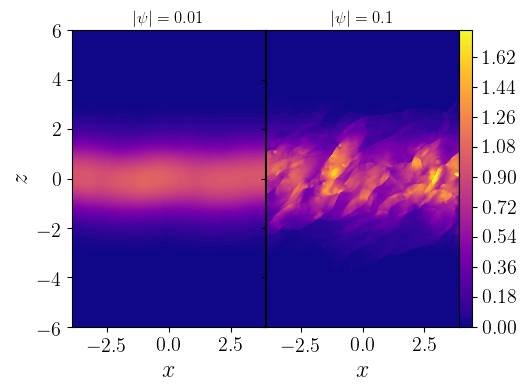}
\caption{Snapshots of the density $\rho$ for an inviscid disc with $q=1.5$ in the nonlinear phase for two different warp amplitudes, both at $\Omega t=200\pi$. Left panel: $\warp=0.01$ for $N=512$; right panel: $\warp=0.1$ for $N=1024$.}
\label{fig:dens_nolaminar}
\end{figure}

The large warp amplitude goes hand in hand with strong density perturbations, as illustrated in figure \ref{fig:dens_nolaminar}. Note that the maximum of the unperturbed density is defined to be unity. While for $\warp=0.01$ perturbations due to the nonlinear inertial waves can be seen, for $\warp=0.1$ density perturbations are much stronger and feature much smaller scales including some weak shocks. This type of flow persists to at least $\warp=0.2$. 


\section{Discussion and conclusions}
\label{secDisc}

We have investigated the hydrodynamic stability of the laminar flows driven by the presence of a warp in the disc. We have confirmed the existence and growth rates of a linear parametric instability \citep{gammie00, OL13b}, both in inviscid non-Keplerian discs and in Keplerian discs for two values of the viscosity parameter, $\alpha=0.01$ and $\alpha=0.001$. At high viscosity, the saturated state consist mostly of wave activity (at nonlinear amplitudes), while towards lower viscosity, a more turbulent state is found. In the Keplerian case, the internal torques $Q_1$ and $Q_2$ were reduced, most severely at low viscosity. Towards larger warp amplitudes, we find that $Q_1>0$, indicating that the mass distribution would evolve in an anti-diffusive manner, possibly breaking the disc up into distinct rings. The reduction in $Q_2$ shows that the warp diffuses on a much longer time scale than one would conclude based on the laminar flows. Since in the saturated state, we no longer have $|Q_2|\gg |Q_3|$ one might expect the warp to evolve in a more wavelike manner.

This hydrodynamic instability to the best of our knowledge has not been seen in global simulations of warped accretion discs. Armed with the results of section \ref{secRes}, we can shed some more light on why this may be the case, focusing on Keplerian discs. First of all, there is a minimum resolution for the instability to develop. While even at our lowest resolution ($\approx 16$ cells per $H$) we found growing modes, at high viscosity the growth rate was underestimated by the low-resolution simulations (see figure \ref{fig:FH_sh15}). Note that this resolution is still three times higher than used in \cite{fragner10}, who used 5 cells per scale height in the meridional direction, while \cite{lodato10} achieved $\sim 3.5$ SPH smoothing lengths per scale height at their highest resolution. For larger warps at low viscosity, the required resolution for capturing the saturated state is more like 128 cells per $H$ at $\warp=0.1$, a resolution that is very difficult to achieve in a global calculation.

The amplitude at which the instability saturates depends on the level of viscosity. At $\alpha=0.01$, the amplitude remains relatively small and the effect on the torque components is modest ($Q_2$ is reduced by a factor of $\sim 2$ at most), while at $\alpha=0.001$ we find larger amplitudes and a strong effect on the torque components ($Q_2$ is reduced by a factor $\sim 100$). If this trend continues towards higher viscosity, the instability may even be difficult to identify for larger values of $\alpha$, which are commonly used in SPH simulations, in combination with a bulk viscosity  \citep[e.g.][]{lodato10}, which we have ignored.

In addition to a resolution constraint, there is the interesting result that no growing modes were found towards larger values of $\warp$, probably due to the strong shear in the laminar flow. At large viscosity ($\alpha=0.01$), only the laminar state is available to the flow for $\warp > 0.07$. It may be that many global calculations are in such a regime. At low viscosity ($\alpha=0.001$), growing modes cease to exist for $\warp > 0.03$. Beyond this warp amplitude, both a laminar and a saturated state are available, and in order to recover the saturated state the initial conditions have to be far enough away from the laminar state. We have found this could be achieved by either starting from a saturated state at lower warp amplitude or by starting from purely Keplerian flow. Note however that the regime $\alpha \lesssim 0.01$ is difficult to achieve with SPH \citep{lodato10}.    

We have considered only the simplest form of the warped shearing box: isothermal and no $y$-dependence. These assumptions should be relaxed in future studies. In particular, a stably stratified disc would give rise to inertia-gravity waves rather than pure inertial waves, while a 3D box would allow for non-axisymmetric shearing waves that may be transiently amplified. Both may have an impact on the growth and saturated state of the hydrodynamic instability.


\section*{Acknowledgements}
SJP is supported by a Royal Society University Research Fellowship. GIO is supported by STFC grant ST/P000673/1.

\bibliographystyle{mnras}
\bibliography{paardekooper}

\bsp
\label{lastpage}

\end{document}